\documentclass[12pt,twoside,a4paper]{article}
\usepackage{graphicx,lipsum,float,enumerate,caption,framed,emptypage}

\usepackage{amsmath}
\usepackage{amsfonts}
\usepackage{amssymb}
\usepackage[scale=0.80]{geometry}
\title{\textbf{A comparison between classical and  Bohmian quantum chaos}}

\author{Athanasios C. Tzemos\footnote{atzemos@academyofathens.gr} \, and George Contopoulos\\ Research Center for Astronomy and Applied Mathematics \\of the  Academy of Athens\\  Soranou Efessiou 4, GR-11527 Athens, Greece }

\date{}

\begin{document}
\maketitle
\begin{center}\textbf{Abstract}\end{center}

We study the emergence of chaos in a 2d system corresponding to a classical Hamiltonian system $V=
\frac{1}{2}(\omega_x^2x^2+\omega_y^2y^2)+\epsilon xy^2$ consisting of two interacting harmonic oscillators and compare the classical and the Bohmian quantum trajectories for increasing values of $\epsilon$. In particular we present an initial quantum state composed of two coherent states in $x$ and $y$, which in the absence of interaction produces ordered trajectories (Lissajous figures) and an initial state which contains {both chaotic and ordered} trajectories for $\epsilon=0$. In both cases we find that, in general, Bohmian trajectories  become chaotic  in the long run, but chaos emerges at times which depend on  the strength of the interaction between the oscillators.

\section{Introduction}\label{sec1}

 In Classical Mechanics (CM) chaos has been well understood and defined as the high sensitivity of the trajectories of bounded nonlinear systems on their initial conditions \cite{Contopoulos200210,wiggins2003introduction}. However, in the framework of Standard Quantum Mechanics (SQM) the evolution of quantum systems is governed by Schr\"{o}dinger's equation which is linear. In addition Heisenbergs uncertainty relation states that it is not possible to find  simultaneously the position and momentum of a quantum particle with arbitrary accuracy. This implies that SQM can not predict trajectories for the quantum particles, contrary to CM. But since macroscopical systems are composed of quantum components it is important to understand  whether chaos exists at the quantum scale.
Thus chaos in Quantum  Mechanics has been an open field of research for several decades. 

Due to the difficulties arising in defining  chaos in SQM most works in the field have been focused on  the behaviour of quantum systems whose classical counterparts have ordered or chaotic Dynamics \cite{haake1991quantum,gutzwiller2013chaos,wimberger2014nonlinear,robnik2016fundamental}, something that has been characterized as `quantum chaology, not quantum chaos' by  Berry in  \cite{berry1989quantum}.


However, Bohmian Quantum Mechanics (BQM), according to which the quantum particles follow well defined deterministic trajectories, similarly to CM, \cite{Bohm,BohmII,holland1995quantum} gives us the opportunity to make a comparison between chaotic dynamics both in the classical and in the  Bohmian quantum framework, based on the standard definition of chaos. This comparison is a fundamental problem that has not been addressed up to now and it is the main subject of the present paper.

Bohmian trajectories are guided by the usual wavefunction, i.e. the solution of the Schr\"{o}dinger equation
\begin{equation}
-\frac{\hbar^2}{2m}\nabla^2\Psi+V\Psi=i\hbar\frac{\partial \Psi}{\partial t},
\end{equation} 
via the Bohmian equations of motion:
\begin{equation}
m\frac{dr}{dt}=\hbar\Im\left(\frac{\nabla \Psi}{\Psi}\right).
\end{equation}
We note that Bohmian Dynamics is non-autonomous and highly nonlinear. Thus when the dimension of the configuration space $N_c$ is $\dim(N_{c})\geq 2$ one observes, in general, the coexistence of ordered and chaotic trajectories \cite{cushing2000bohmian,efthymiopoulos2006chaos,contopoulos2020chaos}. In fact, it has already been shown that there are cases where a classical system is integrable and the corresponding Bohmian quantum system is   chaotic and  there are cases where a chaotic classical system has an ordered Bohmian quantum counterpart.

Most works in Bohmian Mechanics refer to quantum trajectories corresponding to a simple integrable classical system of two or more dimensions, namely a sum of non interacting quantum harmonic oscillators. Considering a 2d system we have 
\begin{equation}
V_0=\frac{1}{2}\omega_x^2x^2+\frac{1}{2}\omega_y^2y^2
\end{equation} and $\Psi$ is, in general, a sum of the energy eigenstates of the form
\begin{equation}\label{lc}
\Psi=\sum_{s}k_{s}\Phi_{m,n}^{(s)},
\end{equation}
where $\Phi_{m,n}^{(s)}=H_m(\bar{x})H_n(\bar{y})$ and $H$ are Hermite polynomials  of degree $m$ in $\bar{x}=\sqrt{\frac{m_x\omega_x}{\hbar}}x$ and $H_n$ of degree $n$ in $\bar{y}=\sqrt{\frac{m_y\omega_y}{\hbar}}y$.{ The indices $m,n$ (quantum numbers) can take  any integer value in $[0,\infty)$.}

A simple case is a product state of only one component of the form:
\begin{equation}
\Psi_{mn}=a\Psi_m(\bar{x})\Psi_n(\bar{y}),
\end{equation}
{which does not have any chaos. In fact this is a product state (it has zero entanglement) and the corresponding Bohmian equations are decoupled, i.e. the $x$ motion is independent from the $y$ motion. Thus its Bohmian trajectories are periodic in the case of commensurable frequences and of Lissajous form in the case of non commensurable frequences.
Then we have the case of two components }
\begin{equation}
\Psi_{mn}=a\Psi_{m_1}(\bar{x})\Psi_{n_1}(\bar{y})+b\Psi_{m_2}(\bar{x})\Psi_{n_2}(\bar{y}),
\end{equation}
{which does not  have chaos as well  since the ratio $dy/dx$ is  time independent and thus it represents a curve on the $x-y$ plane} \cite{tzemos2023order}. 

{The simplest wavefunction which produces chaotic trajectories is that of the superposition of 3 components} \cite{parmenter1995deterministic,makowski2001simplest} {with non commensurable frequencies, i.e.}
\begin{equation}
\Psi_{mn}=a\Psi_{m_1}(\bar{x})\Psi_{n_1}(\bar{y})+b\Psi_{m_2}(\bar{x})\Psi_{n_2}(\bar{y})+c\Psi_{m_3}(\bar{x})\Psi_{n_3}(\bar{y}).
\end{equation}
Therefore, although the classical system $V$ is integrable, the corresponding Bohmian systems in most cases  have both order and chaos. 

Of special importance in Bohmian Dynamics are the nodal points of the wavefunction, where $\Psi_R=\Psi_{Im}=0$ \cite{frisk1997properties,wisniacki2005motion,wisniacki2007vortex}. Trajectories that come close to the nodal points are, in general, chaotic. In fact the flow of particles close to a nodal point in a frame of reference centred at the nodal point has in general a fixed unstable point $X$ and trajectories of particles approaching $X$ are deviated along the unstable asymptotic curves of $X$.  This is the `nodal point-X-point complex mechanism' (NPXPC) \cite{efthymiopoulos2007nodal,efth2009,tzemos2018origin}. The number of nodal points depends on the particular wavefunction $\Psi$. It may be $0, 1, 2, 3\dots$ up to $\infty$. 

We note that the interaction between a Bohmian particle and the X-points of the NPXPC is the main mechanism responsible for the emergence of chaos in its  trajectory. However, as we showed in a previous paper \cite{tzemos2023unstable} there are also unstable fixed points  in the inertial frame, the `Y-points', that also deviate chaotically the approaching particles,  but their contribution is smaller than that of the X-points. With the X-points and the Y-points one can understand  the  time evolution of the deviations of nearby trajectories that lead to the Lyapunov characteristic number.

In our papers we have considered all the above cases with both rational and irrational ratio of the frequencies $\omega_x/\omega_y$. However little Bohmian work has been done in systems with classically chaotic dynamics \cite{sengupta1996quantum,contopoulos2012order}.

Thus it is of interest to consider in a systematic way systems that deviate from the classical harmonic oscillator and calculate the corresponding Bohmian trajectories.

In the present paper we consider a perturbed classical 2d oscillator of the form
\begin{equation}
V=\frac{1}{2}\omega_x^2x^2+\frac{1}{2}\omega_y^2y^2+\epsilon xy^2
\end{equation}
with increasing values of the perturbation parameter $\epsilon$ and we find the corresponding energy spectrum in QM. The perturbation term $\epsilon xy^2$ is not symmetric in $x,y$ and thus it does not produce a symmetric effect on the dynamics of the Bohmian trajectories. 

 In order to make a comparison of the corresponding trajectories in the classical and the quantum framework, we  work with two different wavefunctions in the quantum case: a) a product of 1d coherent states of the quantum harmonic oscillator and b) a superposition of three energy eigenstates. The first one is characterized by a minimum product of the uncertainties in position and momentum and represents the closest quantum system to that of the classical case. In fact if $\epsilon=0$ both the blob of the probability density $|\Psi|^2$  and the corresponding Bohmian trajectories make Lissajous figures, similarly to the classical system. In fact we have studied this wavefunction in great detail in our previous papers on Bohmian qubit systems (see for example \cite{tzemos2019bohmian,tzemos2020ergodicity}).
 
Regarding the second wavefunction, this is the most well studied case in the field of Bohmian chaos, since in the case $\epsilon=0$ it has a single nodal point something that simplifies significantly all the calculations.

We note now that in SQM it is always feasible to proceed with a numerical solution of the Schr\"{o}dinger equation of the perturbed system. However from a Bohmian standpoint it is always better to have an analytic form of the wavefunction. In fact, all calculations involved in the study of Bohmian chaos are based on highly accurate detection of the nodal points and the X-points where chaos is produced. A numerical solution of the Schr\"{o}dinger equation would imply a numerical calculation of the gradients of the wavefunction in the Bohmian equations of motion before their (numerical also) integration. This process is prone to numerical errors and time consuming even for a powerful computer, especially when one wants to integrate the Bohmian trajectories for very large times in order to observe long time effects as e.g. ergodicity \cite{tzemos2020ergodicity}.

Therefore in the present work we choose to solve the Schr\"{o}dinger equation  (SE) by diagonalizing the perturbed Hamiltonian. In order to do so, we consider a set of the energy levels of the unperturbed ($\epsilon=0$) 2d quantum harmonic oscillator and exploit the fact that its corresponding eigenstates form a complete orthonormal basis $|\Phi_m\rangle$ in Hilbert's space, where $|\Phi_1\rangle=|0\rangle_x\otimes|0\rangle_y\equiv |00\rangle$, $|\Phi_2\rangle=|01\rangle$,$|\Phi_3\rangle=|10\rangle$, $|\Phi_4\rangle=|02\rangle$,$|\Phi_5\rangle=|11\rangle$, $|\Phi_5\rangle=|20\rangle\dots$ in the Hilbert space, where $|0\rangle, |1\rangle, |2\rangle,\dots$ refer to the energy eigenstates of the corresponding quantum numbers $n_x$ and $n_y$ in $x$ and $y$. Thus we write the Hamiltonian operator $\hat{H}$ as a matrix $H_{q,r}$ in the basis  of $|\Phi_m\rangle$ where 
\begin{equation}
H_{q,r}=\langle \Phi_q |\hat{H} |\Phi_r \rangle.
\end{equation}
In the position representation an element of the above matrix reads:
\begin{equation}
H_{q,r}=\int_{-\infty}^{\infty}\Phi_q^{\dagger}(H_0+\epsilon xy^2) \Phi_r dxdy,
\end{equation}
where $\Phi_m(x,y)$ is the product of the 1d wavefunctions of the energy levels $n_x,n_y$ inside $\Phi_m$.
The unperturbed part $\mathbf{\hat{H}_0=\left(\frac{\hat{p}_x^
2}{2m_x}+\frac{\hat{p}_y^2}{2m_y}\right)+V_0}$ will produce the diagonal elements of the above matrix which for given $n_x$ and $n_y$ in the mth basis state $\Phi_m$ which correspond to the energy levels
\begin{eqnarray}\label{en}
E_m(n_x,n_y)=(n_x+\frac{1}{2})\hbar\omega_x+(n_y+\frac{1}{2})\hbar\omega_y.
\end{eqnarray}
{The perturbation will shift the energy levels of the unperturbed system (the diagonal terms of the Hamiltonian) and produce as well the non diagonal elements of $H_{q,r}$ which indicate coupling between different energy levels.}
In our calculations we take $\omega_x=1, \omega_y=\sqrt{2}/2$. The condition $\omega_x/\omega_y=$irrational is  necessary but not sufficient for the existence of chaos in Bohmian trajectories.

The diagonalization of the above matrix can be easily made with a computer algebra system (we work here with MAPLE), when all the numerical values of the physical constants are given in advance. Then the matrix is purely numerical and the eigenvalue problem is solved by the computer, which finds the eigenvalues $\tilde{E}_n$  and the corresponding  eigenstates $|\tilde{\Phi}_n\rangle$ of the perturbed system which can be written in the form: \cite{merzbacher1998quantum,ballentine2014quantum}
\begin{equation}
\tilde{|\Phi_n}\rangle=\sum_m c_{mn}|\Phi_m\rangle
\end{equation}

Thus for a given initial state $|\Psi_0\rangle$  we find
\begin{equation}
|\Psi(t)\rangle =\sum_n e^{-i\frac{\tilde{E}_nt}{\hbar}}\langle \tilde{\Phi}_n|\Psi_0\rangle|\tilde{\Phi}_n\rangle\\
=\sum_{m,m',n}c_{m'n}^{*}c_{mn}e^{-i\frac{\tilde{E}_nt}{\hbar}}\langle \Phi_{m'}|\Psi_0\rangle|\Phi_m\rangle,
\end{equation}
with $m,m',n$ running from 1 up to $N$, where $N$ is the number of different energy levels that we consider in our truncation.
Thus the state of the full system is written as a superposition of the states $|\Phi_m\rangle$ of the unperturbed system.

We note that the choice of the initial state $|\Psi_0\rangle$ will affect the number of the energy levels that one should take into account. The higher their number the better the approximation\footnote{ In any case with a computer algebra system it is easy to diagonalize very quickly numerical Hamiltonian matrices with dimension  larger than $1000\times 1000$.}. However, we want to study Bohmian trajectories and if we take a great number of energy levels in our Hamiltonian matrix, then we have to deal with complicated derivatives over many Hermite polynomials which result to extremely long expressions for the Bohmian equations, something that increases dramatically  their integration time (see the Appendix).

In the present paper we consider in section 2 an initial state composed of two 1d coherent states in $x$ and $y$, truncated at a high order of the sum of the corresponding quantum numbers $n=n_x+n_y$. This case does not have any chaos for $\epsilon=0$ but it is completely chaotic for $\epsilon\neq 0$. Then in Section 3 we consider a quantum case that has some chaos, even for $\epsilon=0$, and find its chaos for $\epsilon\neq 0$. In Section 3 we consider the corresponding classical case and in section 5 we compare the various cases by emphasizing the similarities and differences between the classical trajectories and the Bohmian quantum trajectories and draw our conclusions. {Finally in Appendix A we briefly describe our the diagonalization process that we followed while in Appendix B we describe the practical method for distinguishing between order and chaotic trajectories providing two examples of Lyapunov exponents.}

\section{An initial coherent state with no chaos at $\epsilon=0$}

We now work with a wavefunction which is a product of two coherent states of the quantum harmonic oscillator in $x$ and $y$ coordinates. The coherent states are a special class of solutions of the SE. A coherent state $|\alpha\rangle$ is defined as the eigenstate of the annihilation operator $\hat{a}=\sqrt{\frac{m_x\omega_x}{2\hbar}}\left(\hat{x}+\frac{i}{m_x\omega_x}\hat{p}\right )$ \cite{garrison2008quantum}: 
\begin{align}
\hat{a}|\alpha\rangle=A|\alpha\rangle,
\end{align}
where the eigenvalue $A$ is, in general, a complex number since $\hat{a}$ is not Hermitian. Namely, $A=|A|\exp(i\theta)$, where $|A|$ is the amplitude and $\theta$ the phase of the state $|\alpha\rangle$. 
In the position representation the time evolution of a {1-D coherent state} reads 
\begin{align}\label{coh}
Y(x,t)=e^{-\frac{1}{2}a_0^2}e^{\frac{-i\omega_x t}{2}}\sum_{n=0}^{n_f}\frac{(a_0e^{i\sigma_x} e^{-i\omega_x t})^n}{\sqrt{n!}}\psi_n(x),
\end{align}
where 
\begin{align}\label{coh2}
\psi_n(x)=\frac{1}{\sqrt{2^nn!}}\left(\frac{m_x\omega_x}{\pi\hbar}\right)^{\frac{1}{4}}e^{-\frac{m_x\omega_x x^2}{2\hbar}}H_n\left(\sqrt{\frac{m_x\omega_ x}{\hbar}}x\right),n=0, 1, 2,\dots,
\end{align}
{and}
$H_n(k)=(-1)^ne^{k^2}\frac{d^n}{dk^n}\left(e^{-k^2}\right)$ {are the corresponding Hermite polynomials}.
For a complete coherent state we have $n_f=\infty$. Moreover
 $a_0=|A(0)|,  \sigma_x, \omega_x$ are the initial values of the amplitude, the initial phase, and the frequency of the oscillator.
Although there are infinitely many energy eigenstates in  a coherent state, their spectrum follows Poissonian statistics i.e. the probability  $P_S(n)$ {of detecting energy level} $n$ in the coherent state $|\alpha\rangle$ is \cite{garrison2008quantum}
\begin{align}
P_S(n)=|\langle n|\alpha\rangle|^2=\frac{e^{-\langle n\rangle}\langle n\rangle ^n}{n!}.
\end{align} 

\begin{figure}[H]
\centering
\includegraphics[width=0.45\textwidth]{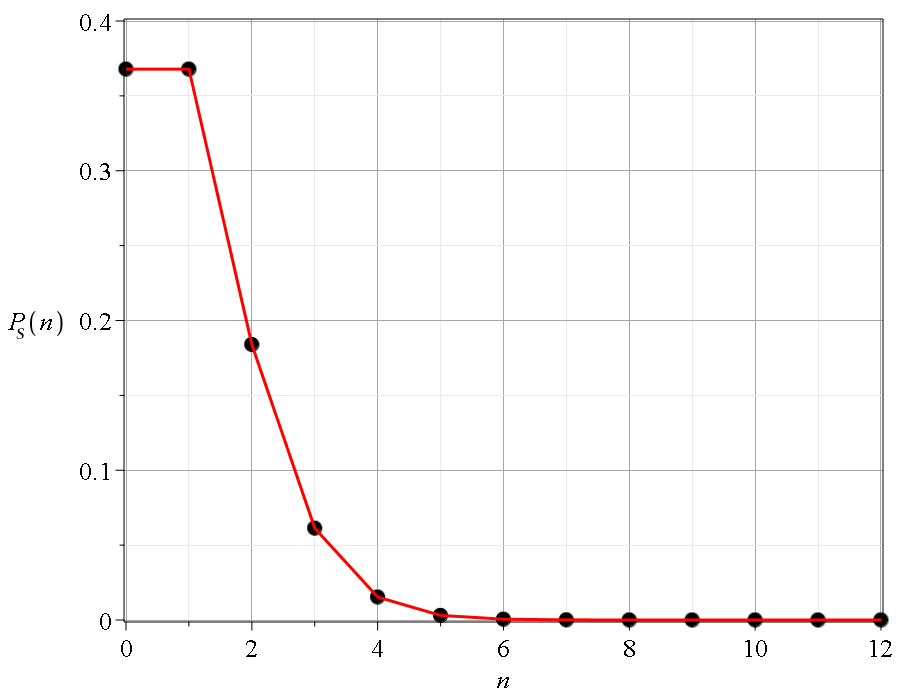}
\caption{a){ The Poisson distribution of the energy levels inside each of the 1d coherent states along $x$ and $y$ forming $\Psi_0=Y(x,0)\cdot Y(y,0)$  with common $a_0=1$  up to $n=12$}. We find that $P_S(6)\simeq 5.1\times 10^{-4}$ and $P_S(12)\simeq 7.7\times 10^{-10}$.}\label{poisson}
\end{figure}

We note that the mean value $\langle n\rangle$ and the variance $(\Delta n)^2$  are both constant, equal to $|A|^2$, and  $\sum_{n=0}^{n=\infty}P_S(n)=1$  (see Fig.~\ref{poisson}). Thus it  is always possible to approximate efficiently a coherent state if we consider a sufficient number of energy levels (we see that $P_S$ is very small for $\epsilon\geq 0.6$ for our parameters).

\begin{figure}[H]
\centering
\includegraphics[width=0.45\textwidth]{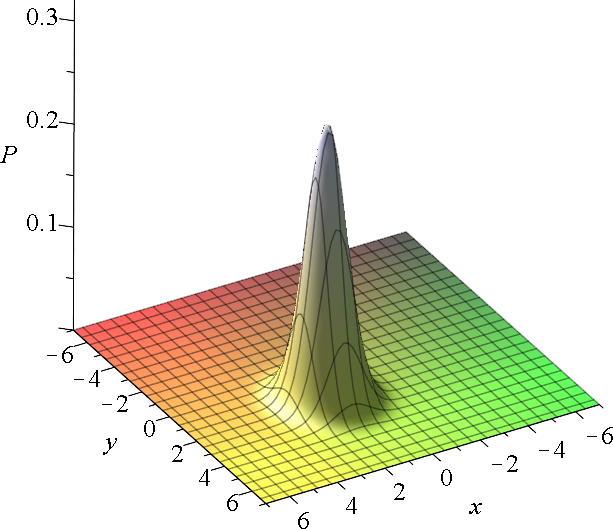}
\caption{The probability density $P(x,y)=|\Psi(x,y)|^2$ of the product of two coherent states in $x$ and $y$ at $t=0$.}\label{poisson2}
\end{figure}

{In the present case we take as an initial state a product of coherent states along $x$ and $y$  axes, i.e. 
\begin{equation}
\Psi_0=Y(x,0)Y(y,0)
\end{equation} with common $a_0=1$, $\sigma_x=\sigma_y=0,\omega_x=1, \omega_y=\sqrt{2}/2$ and $m_x=m_y=\hbar=1$.  According to Born's rule, the probability density of finding a particle at a certain region of space is $P(x,y)=|\Psi(x,y)|^2$. In    Fig.~\ref{poisson2} we give $P_0=|\Psi(x,y;t=0)|^2$ of this product state. The Gaussian blob makes a Lissajous figure in the case of $\epsilon=0$ \cite{tzemos2019bohmian,tzemos2020ergodicity}.}

In order to limit the energy levels in our Hamiltonian we use two coherent states in $x$ and $y$ with common amplitude $a_0=1$ so that the maximum contribution is  at  $n=1$ (Fig.~\ref{poisson}a). Thus we take into account all the possible 91 doublets $(n_x,n_y)$ whose sum is  up to $n_x+n_y=12$.  Namely we write the wavefunction of the perturbed system as a superposition of the $|\Phi_m\rangle$ states of the unperturbed system:
\begin{align}
|\Psi(t)\rangle=\sum_{m=1}^{91} q_m(t)|\Phi_m\rangle,
\end{align}
where
\begin{align}
q_m(t)=\sum_{m',n}c_{m'n}^{*}c_{mn}e^{-i\frac{E_nt}{\hbar}}\langle \Phi_{m'}|\Psi_0\rangle.
\end{align}
The average energy  $E_{av}$ is also a time dependent quantity
\begin{equation}\label{eaveq}
E_{av}=\sum_{m=1}^{91}P_m(t)E_m,\quad \text{with} \, P_m=|q_m(t)|^2.
\end{equation}

\begin{figure}[H]
\centering
\includegraphics[width=0.27\textwidth]{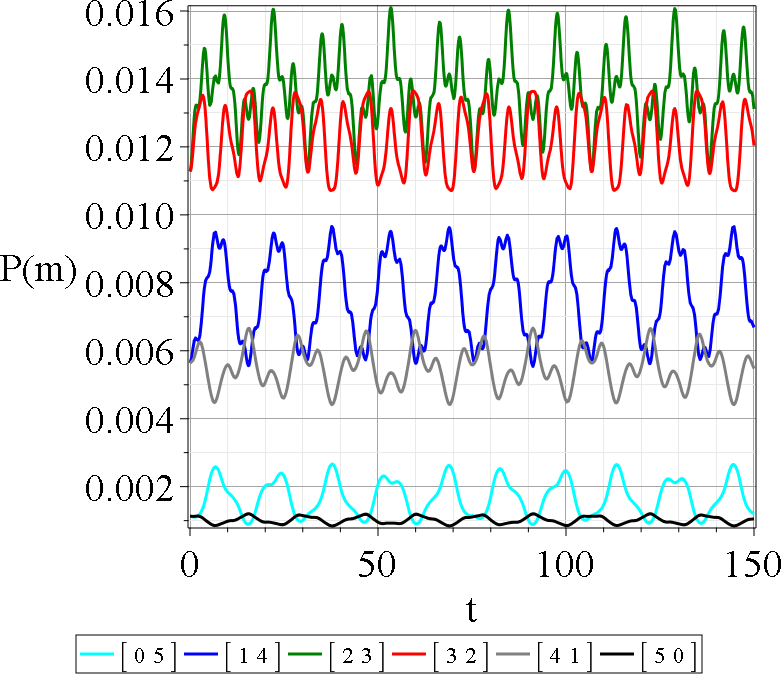}[a]
\includegraphics[width=0.27\textwidth]{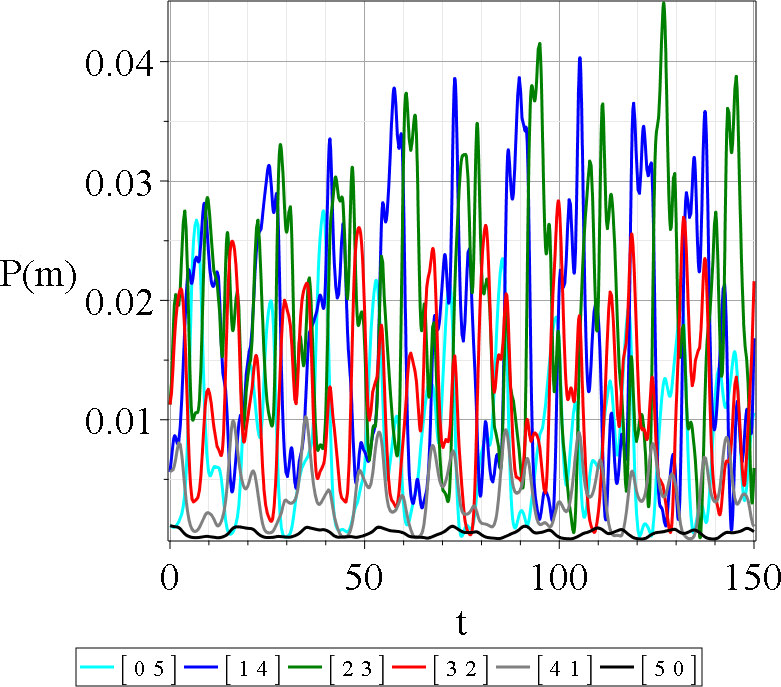}[b]
\includegraphics[width=0.27\textwidth]{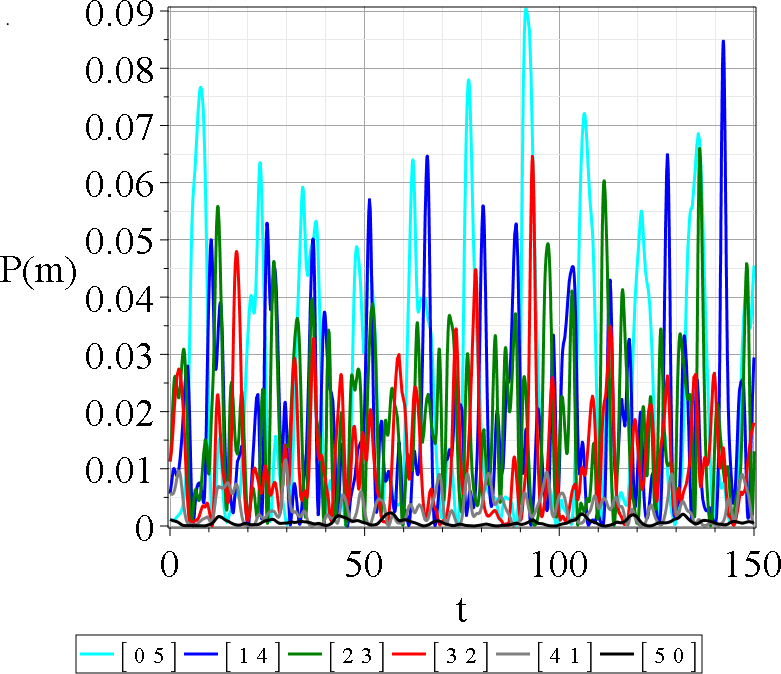}[c]
\includegraphics[width=0.27\textwidth]{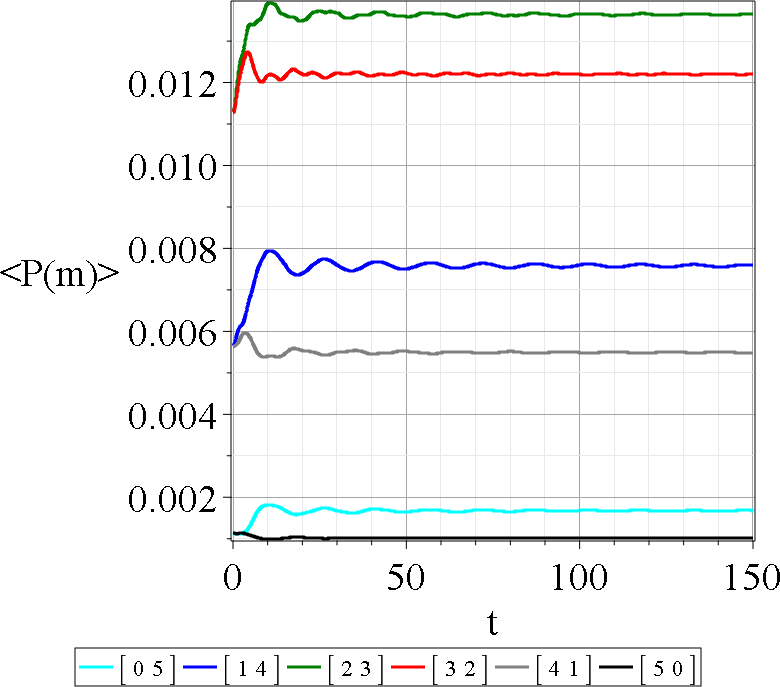}[d]
\includegraphics[width=0.27\textwidth]{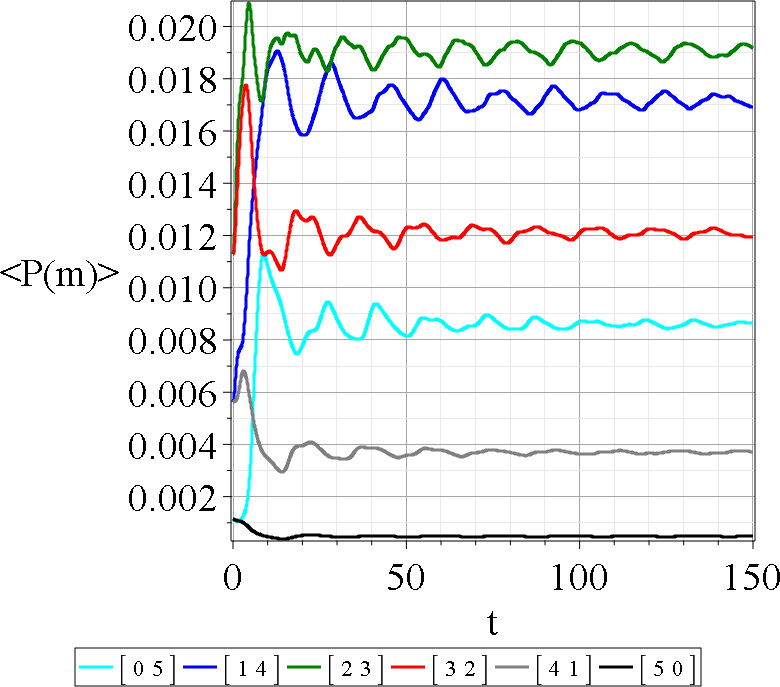}[e]
\includegraphics[width=0.27\textwidth]{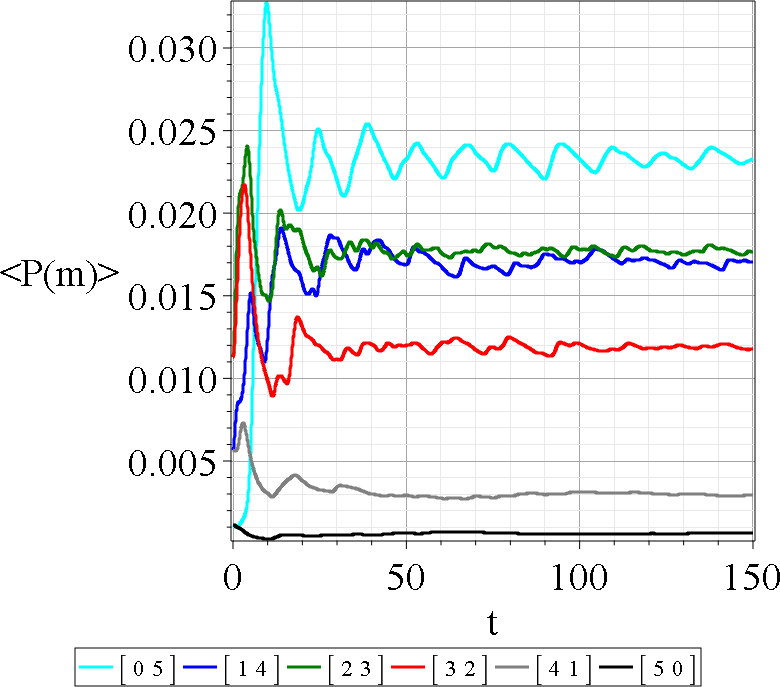}[f]
\caption{Upper panel: The  time evolution of the probability $P(m)$ of the energy levels with $n_x+n_y=5$ for $\epsilon=0.01$ (a), $\epsilon=0.05$ (b) and $\epsilon=0.09$ (c).
Lower panel: The corresponding mean value $\langle P(m)\rangle$ up to time $t$. We observe that as the perturbation parameter $\epsilon$ increases the contribution of the higher energy levels increases, in general, as well. Similar figures exist for larger $\epsilon$. However, for all the values of $\epsilon$ considered in the present paper our truncation of the energy levels gives sufficiently accurate results. The colors are upwards: black=[5,0], cyan=[0,5], grey=[4,1], blue=[1,4], red=[3,2] and green=[2,3].}\label{timeseries}
\end{figure}

In Fig.~\ref{timeseries} we show the probability of finding the perturbed system in the mth eigenstate of the unperturbed system. In the upper panel we plot the time evolution of $P(m)$, where $m$ refers to the different energy levels whose quantum numbers $n_x,n_y$ sum up to 5  for $\epsilon= 0.01, 0.05$ and $0.09$ respectively. In the lower panel we show the corresponding time evolution of the average $\langle {P}(m)\rangle$, up to time $t$, for the same energy levels. The mean values have some oscillations, but they tend to stabilize with the increase of time. {Similar figures are found for sums of quantum numbers $n_x+n_y$, equal to $1, 2, \dots$ up to 12. We also observe that  the contribution of the higher energy levels is larger when the interaction term $\epsilon$ increases but  not very much.  }

{Therefore  for larger $\epsilon$ we need a larger truncation. However, in our case where we study a wavefunction of low energy, the contribution (i.e. the probabilities) of the higher energy levels of our truncation is very small. {In particular the mean probability of the last energy level states ($n_x+n_y=12$) is found to be of the order $10^{-6}$ for $\epsilon=0.01$, $10^{-5}$ for $\epsilon=0.05$ and $10^{-4}$ for $\epsilon=0.09$.}}

\begin{figure}[H]
\centering
\includegraphics[width=0.4\textwidth]{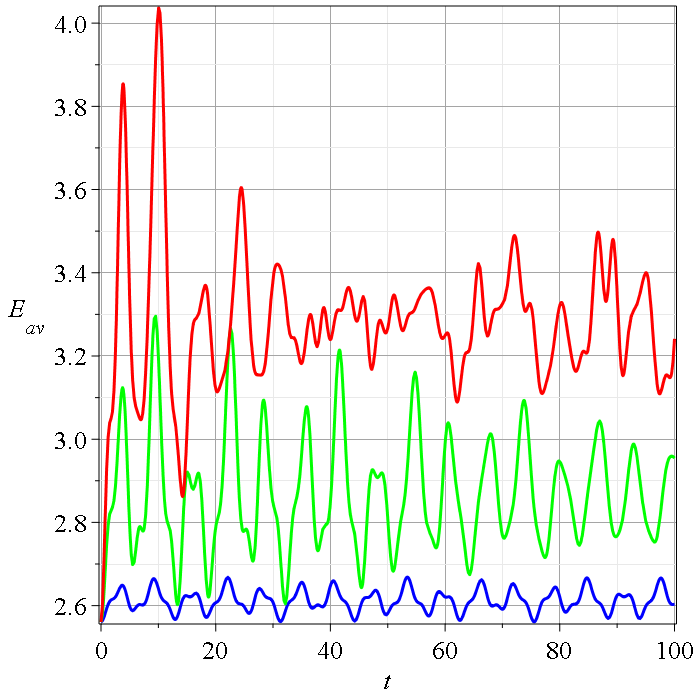}[a]
\includegraphics[scale=0.2]{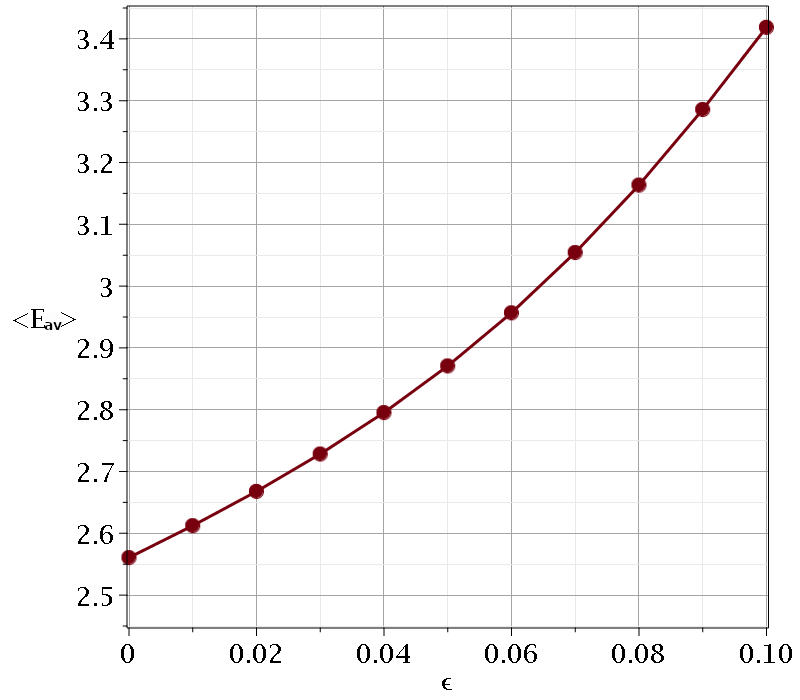}[b]
\caption{The time evolution of the $E_{av}$ for $\epsilon=0.01$ (blue) $\epsilon=0.05$ (green) and $\epsilon=0.09$ (red). The larger the value of the parameter $\epsilon$ the stronger its influence on the interaction between the different energy levels. b) The mean energy of the system for $t\in[0,10^4]$ as a function of the perturbation parameter $\epsilon$.}\label{eav}
\end{figure}

{In Fig.~\ref{eav}a we plot the time evolution of the average energy $E_{av}$ (Eq.~\ref{eaveq}) of our system for $\epsilon=0.01$ (blue curve), $\epsilon=0.05$ (green curve) and $\epsilon=0.09$ (red curve), while in Fig.~\ref{eav}b we plot the corresponding time average of $E_{av}$ as a function of $\epsilon$. Both figures show that the larger the $\epsilon$ the larger the contribution of the higher energy levels in the system.  We note that the final point of Fig.~\ref{eav}b   is $\epsilon=0.097$ , since for $\epsilon\geq 0.097$ most trajectories of the classical system escape to infinity (Section 4) and cannot be compared with the corresponding Bohmian quantum trajectories. 
}

%

\begin{figure}[H]
\centering
\includegraphics[width=0.28\textwidth]{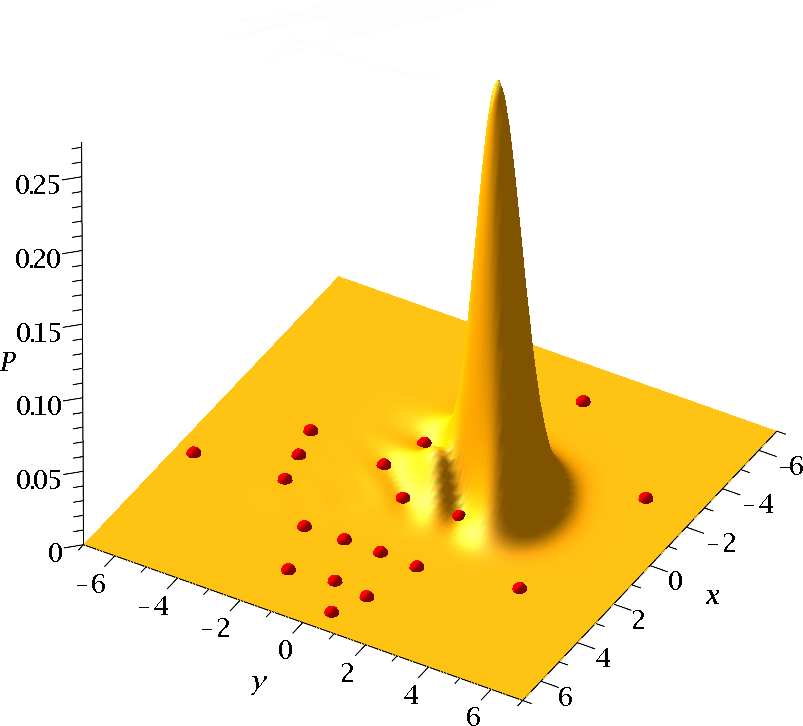}[a]
\includegraphics[width=0.24\textwidth]{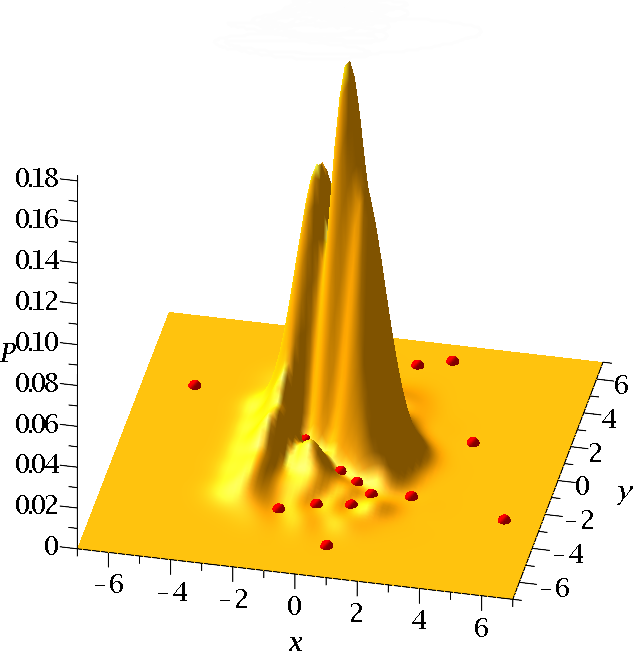}[b]
\includegraphics[width=0.24\textwidth]{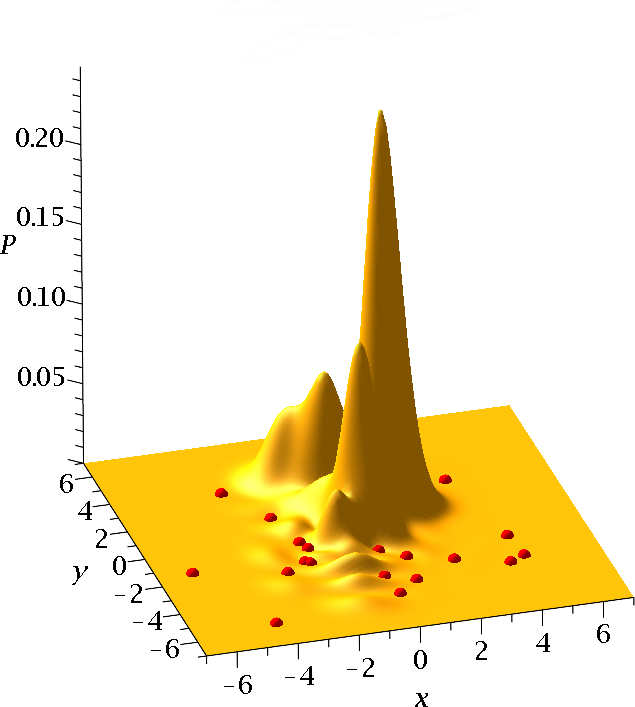}[c]
\caption{ Examples of major deformations of of $P=|\Psi|^2$ for:  a) $\epsilon=0.01$ ($t=1725$), b) $\epsilon=0.05$ ($t=100$) and  c) $\epsilon=0.09$ ($t=20$). The larger the perturbation the quicker the deformation of the shape of $P$ and thus the emergence of chaos.}\label{snaps}
\end{figure}

For $\epsilon=0$ there are no nodal points (they are at infinity). The probability distribution in the Born case ($P=|\Psi|^2$) is initially ($t=0$) a unique blob of the form $|\Psi|^2=|Y(x,0)^2\cdot Y(y,0)|^2$ (Fig.~\ref{poisson}b). As the time increases the interaction of the various levels (Eqs.~\eqref{coh}-\eqref{coh2}) produces a change of the blob. After a certain time the nodal points enter the central region of the blob (Fig.~\ref{snaps}), and then the blob splits or it is significantly deformed  and chaos becomes important.

Chaos is introduced when the quantum particles approach the region close to one of the nodal points. Near every nodal point there is an unstable fixed point $X$ of the flow in the frame of reference of the nodal point. When particles approach it they are deflected chaotically along the opposite directions of the unstable asymptotic curves of this point.

\begin{figure}[H]
\centering
\includegraphics[width=0.29\textwidth]{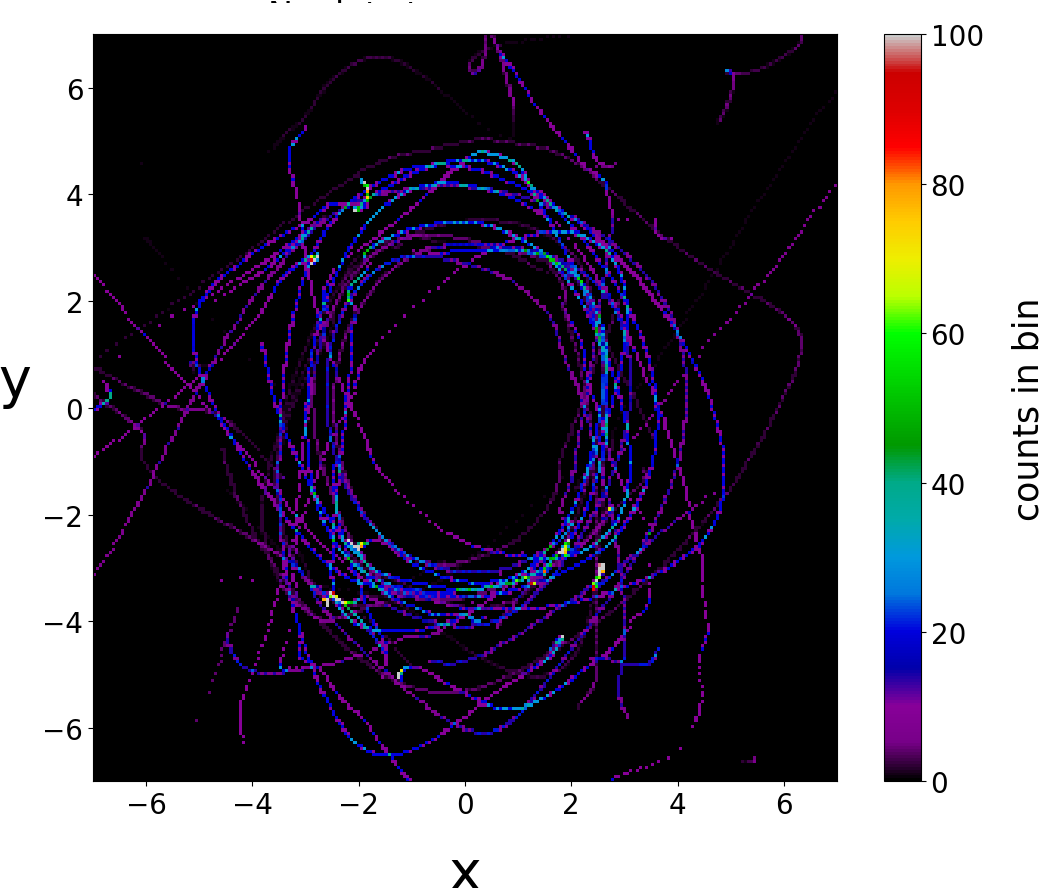}[a]
\includegraphics[width=0.29\textwidth]{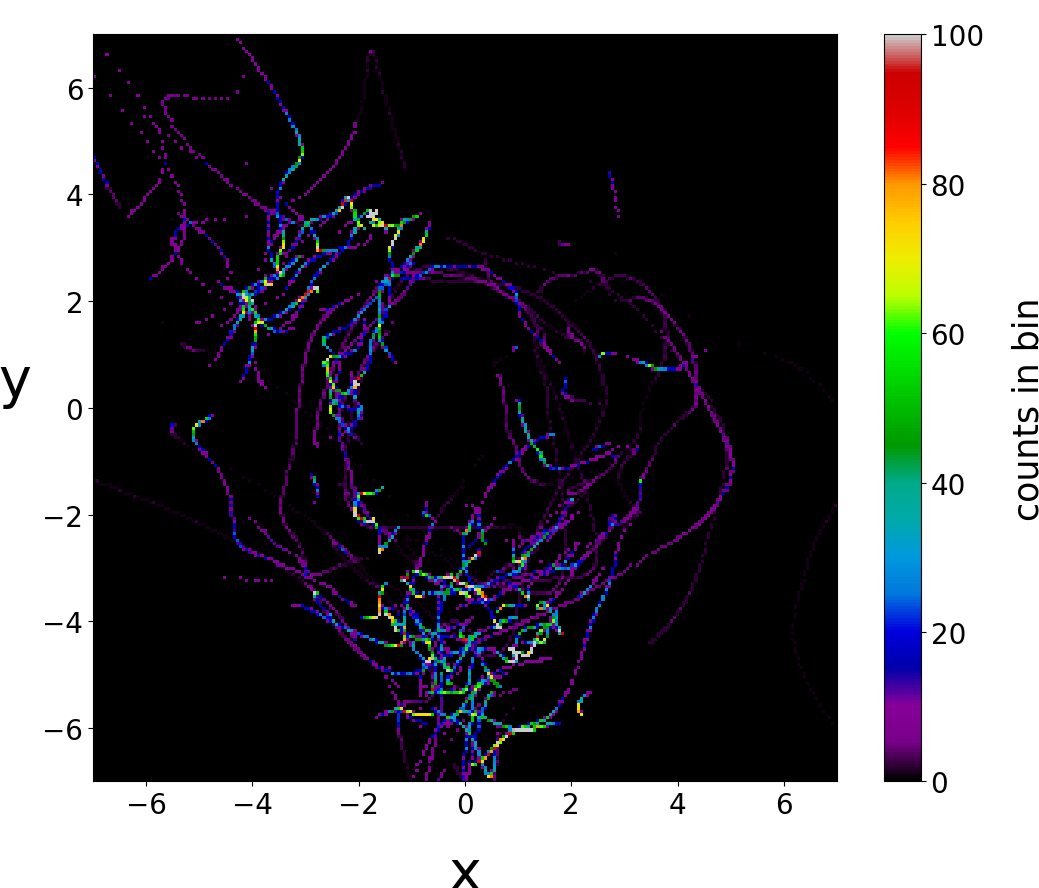}[b]
\includegraphics[width=0.29\textwidth]{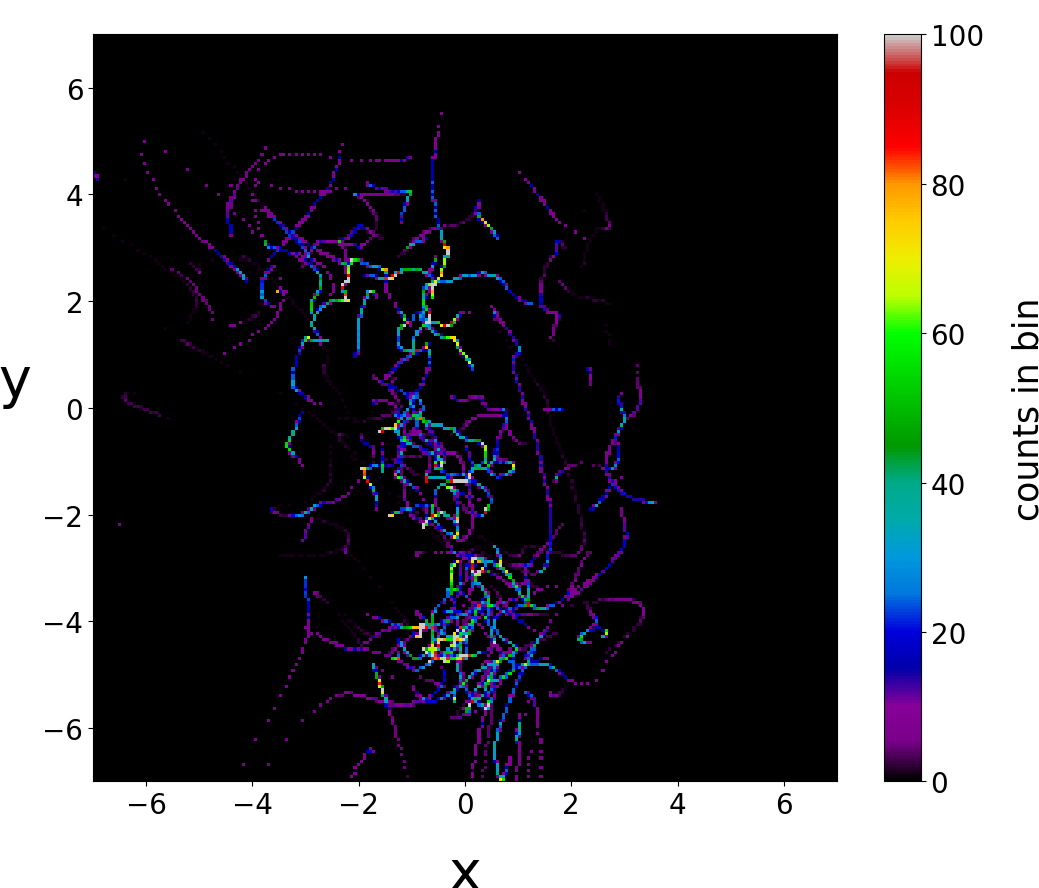}[c]
\caption{The colorplots of the nodal points for $t\in[0,100]$ for the cases: a) $\epsilon=0.01$, b) $\epsilon=0.05$ and c) $\epsilon=0.09$. {The larger the $\epsilon$ the quicker the penetration of the nodal points in the area close to the origin where the probability density has significant values. This implies the quick deformation of the $|\Psi(x,y)|^2$ and the emergence of chaos.}}\label{cp}
\end{figure}

In Fig.~\ref{cp} we give the colorplots of the nodal points (i.e. the number of times that each one of the nodal points enters in  every square bin ($0.05\times 0.05$) of the configuration space up to a fixed time $t=100$.
When $\epsilon$ is small ($\epsilon=0.01$ in Fig.~\ref{cp}a) the nodal points have not yet reached the central region but they are mainly concentrated in a ring around the center. Later on, the nodal points do reach the center as seen inf Fig.~\ref{snaps}a. When $\epsilon=0.05$ at the same time the nodal points approach the center (Fig.~\ref{cp}b) and when $\epsilon=0.09$ (Fig.~\ref{cp}c) they are concentrated near the center.

\begin{figure}[H]
\centering
\includegraphics[width=0.4\textwidth]{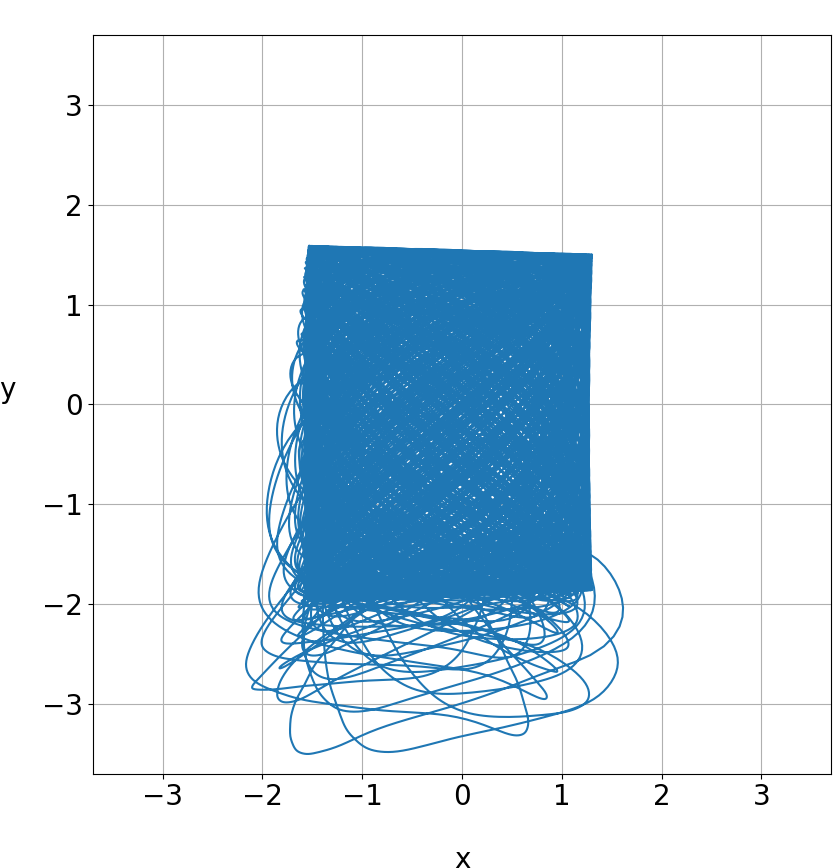}[a]
\includegraphics[width=0.4\textwidth]{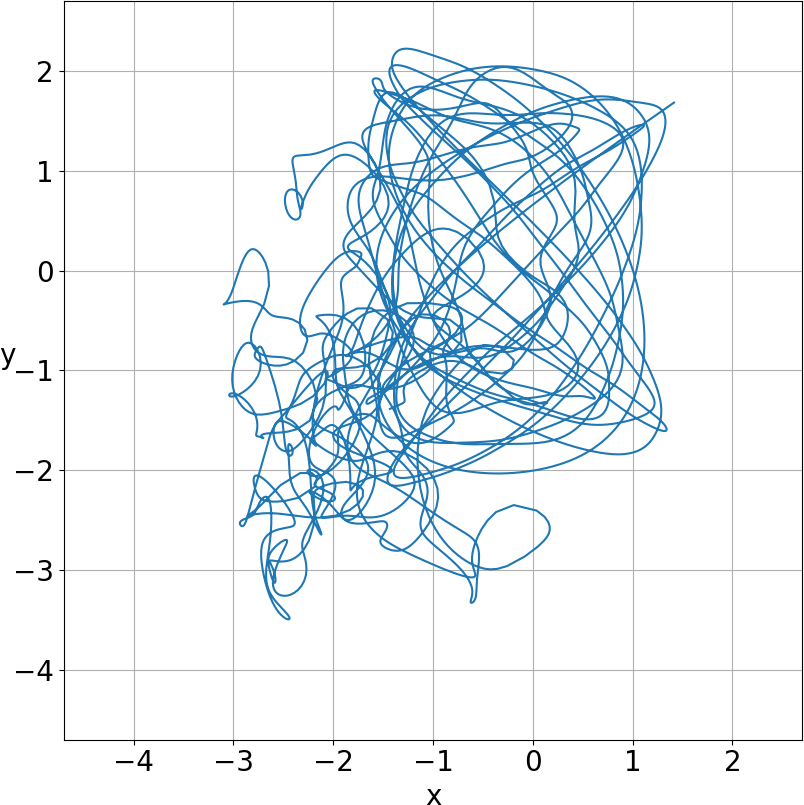}[b]
\caption{Bohmian trajectories for: a)  $\epsilon=0.01$  ($x(0)=1.3, y(0)=1.5$) up to $t=2000$ and b) $\epsilon=0.05$ ($x(0)=1.415, y(0)=1.6826$) up to $t=250$.}\label{traj}
\end{figure}

{The trajectories of most particles have the form of a distorted Lissajous figures (Fig.~\ref{traj}a) as long as they are far away from the nodal points $N$ and their associated X-points. }However, later on the $N$ and X-points come close to the center and  they produce large deformations of the blob. Then the trajectories are deflected and become chaotic (Fig.~\ref{traj}a). When $\epsilon$ is larger the deflections and the chaotic character of the trajectories appears earlier (Fig.~\ref{traj}b).

\begin{figure}[H]
\centering
\includegraphics[width=0.45\textwidth]{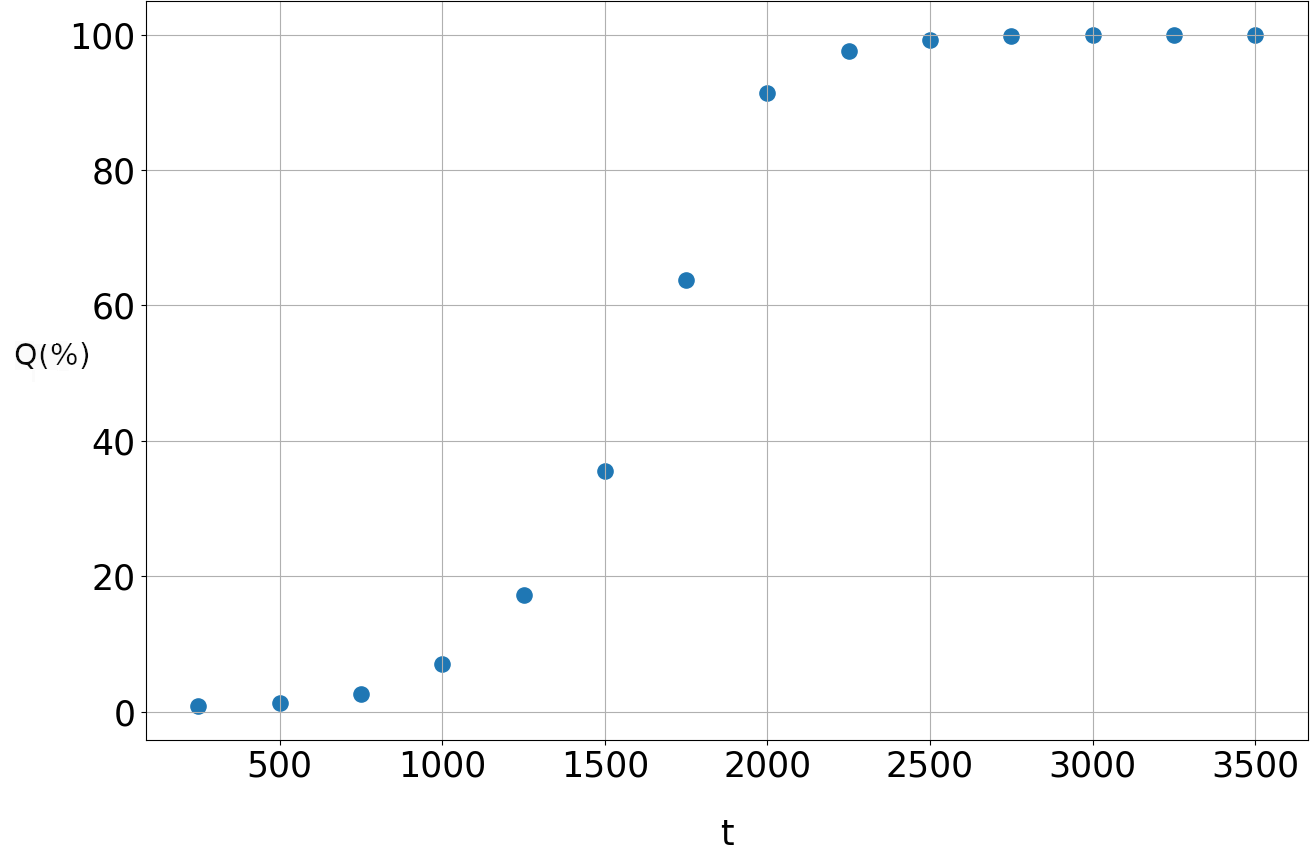}[a]
\includegraphics[width=0.45\textwidth]{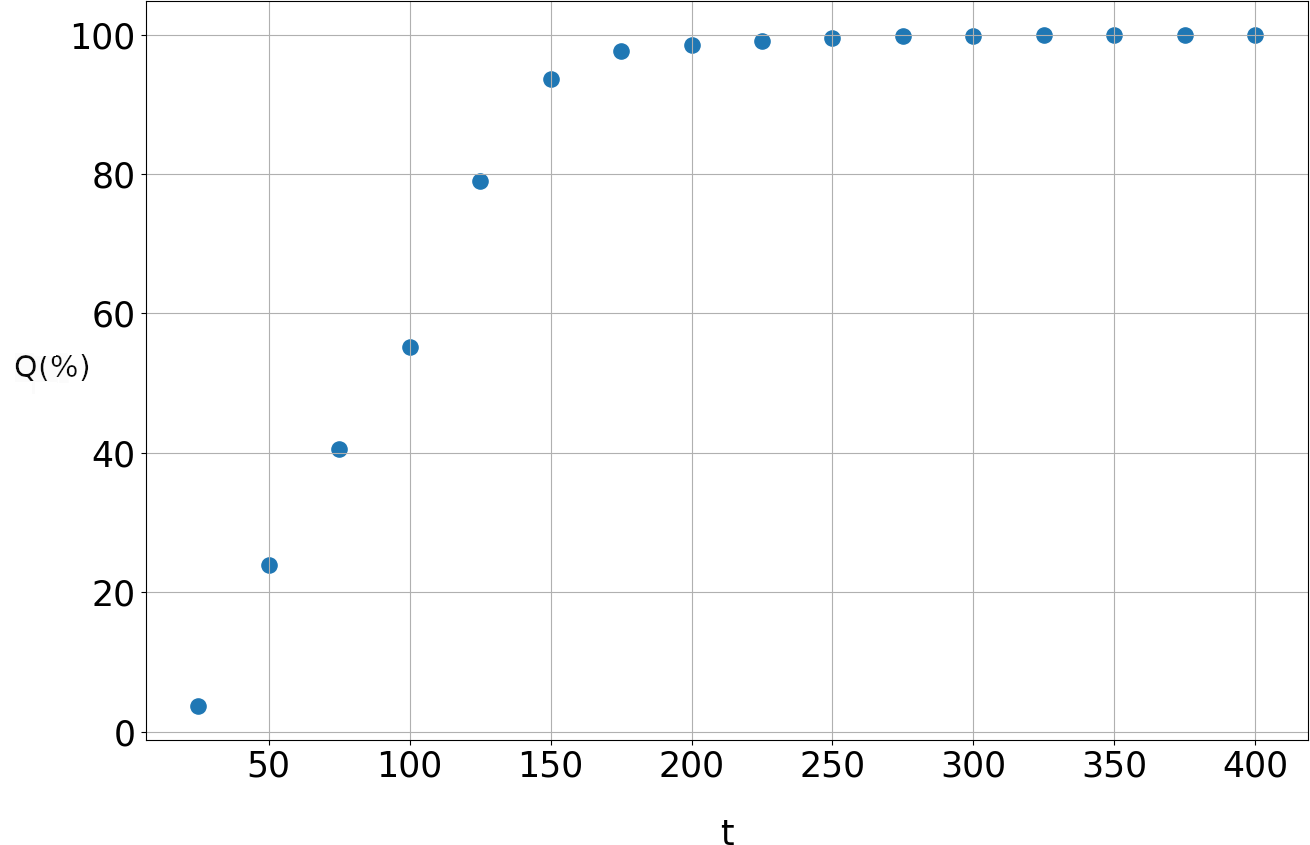}[b]
\caption{a) {The proportion of trajectories found to aquire chaotic behaviour inside a realization of Born's rule with 5000 particles as a function of time} for $\epsilon=0.01$. We observe that at about $t=1500$ we have an inflection point and beyond $t=2500$ the proportion of chaotic trajectories reaches about $100\%$ . b) The same happens in the case $\epsilon=0.05$ but for much shorter times. For $t>200$ practically all the trajectories are chaotic.}\label{br_time}
\end{figure}

In Figs.~\ref{br_time}a we give the proportion of the trajectories that look chaotic as a function of time, for $\epsilon=0.01$. Initially  all the trajectories look like ordered and chaos appears in a small number of them. The number of trajectories found to be chaotic increases gradually, but around $t=1500$ the increase is abrupt. Around this time the blob of the probability density $P$ (Fig.~\ref{snaps}a) starts to have large deformations. Then many trajectories get scattered and chaos increases considerably, reaching about $100\%$ for $t\geq 2500$. In Fig.~\ref{br_time}b  ($\epsilon=0.05$) we see that the increase of the chaotic trajectories becomes  much faster.

\begin{figure}
\centering
\includegraphics[scale=0.2]{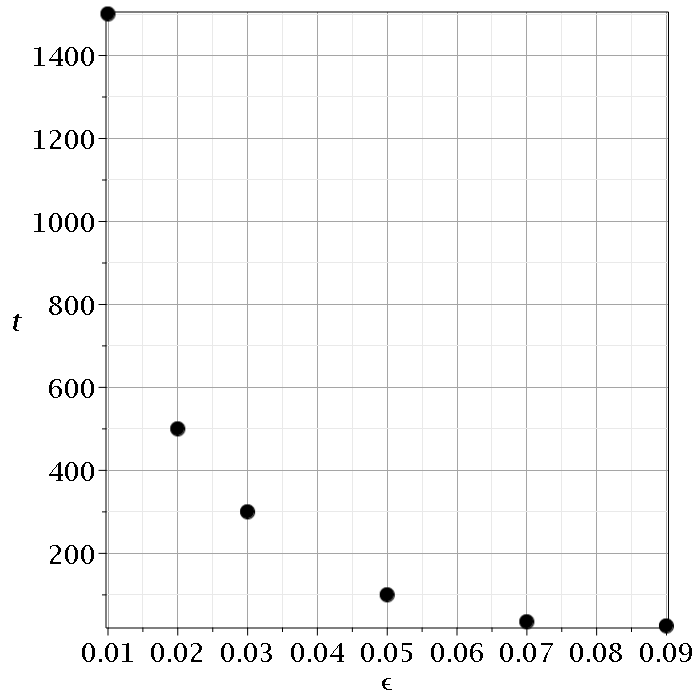}
\caption{An approximate estimation of the time $t_c$ as a function of $\epsilon$ when the nodal points pass through the blob of $|\Psi|^2$ and deflect most of the trajectories, producing a large degree of chaos.}\label{time_vs_e}
\end{figure}

The times of these significant deformations of the trajectories that  characterize them as chaotic are given in Fig.~\ref{time_vs_e}. We see that they are large for small $\epsilon$ (e.g. $\epsilon=0.01$) and decrease drastically as $\epsilon$ increases. In fact for $\epsilon=0.09$ and for larger $\epsilon$ the deformation time is very close to zero. On the other hand as $\epsilon$ decreases and tends to zero the deformation time tends to infinity and for $\epsilon=0$ there is no deformation at all (coherent blob of Fig.~\ref{poisson}b and no chaos) and all trajectories are ordered.

If we fix the time up to which we calculate the trajectories in Fig.~\ref{time_vs_e} we find a transition value at $\epsilon=\epsilon_c$. For $\epsilon<\epsilon_c$ most trajectories look as ordered (close to Lissajous figures) and for $\epsilon>\epsilon_c$ most trajectories appear as chaotic. Therefore the value of $\epsilon_c$ is an effective transition perturbation, similar to the transition perturbations of Classical Mechanics \cite{Contopoulos200210}.

As a conclusion we see that chaos is introduced for practically all the trajectories for $\epsilon\neq 0$. It is remarkable that when $\epsilon\neq 0$ there seem to be no ordered trajectories, but only time differences for the introduction of chaos.

%

\section{A wavefunction with a single nodal point}

A simple wavefunction with only one nodal point for $\epsilon=0$ is 
\begin{equation}\label{wfsn}
\Psi=a\Psi_{0,0}+b\Psi_{1,0}+c\Psi_{1,1}.
\end{equation}
This case has been studied extensively in the past \cite{parmenter1995deterministic,makowski2001simplest,
efthymiopoulos2007nodal} and its main difference
from the coherent states and their approximations considered in section 2, is that it has a proportion (about $5\%$) of chaotic trajectories for $\epsilon=0$. Similar systems corresponding to a classical potential of two harmonic oscillators have also, in general, a proportion of chaos for $\epsilon=0$ \cite{tzemos2023unstable}.

\begin{figure}[H]
\centering
\includegraphics[scale=0.18]{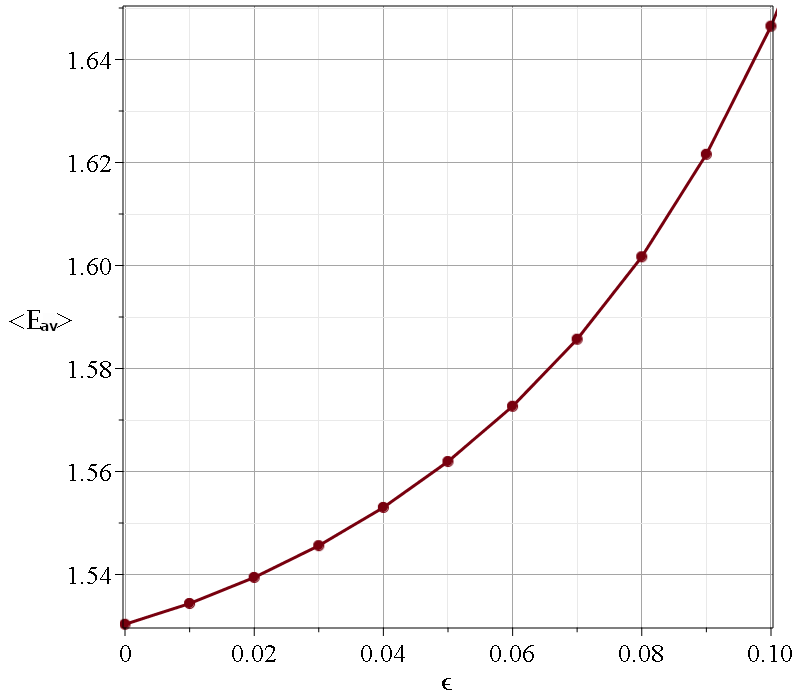}
\caption{The average energy $\langle E_{av}\rangle$ of the wavefunction with a single nodal point as a function of the perturbation $\epsilon$.}\label{ensingle}
\end{figure}

We consider a quantum system with $\omega_x=1,\omega_y=\sqrt{2}/2$ and $a=\frac{1}{\sqrt{2}}, b=c=\frac{1}{2}$, so that $|a|^2+|b|^2+|c|^2=1$. For $\epsilon=0$ the energy levels of the wavefunction are $E_{nx,ny}=n_x+n_y+1$ according to Eq.~\ref{en}. Thus the total energy is $E=a^2E_{00}+b^2E_{10}+c^2E_{11}=1.53$ (Fig.~\ref{ensingle}). In this case the classical system has no chaos at all. The main question now is to find how much quantum chaos appears when $\epsilon$ increases from zero. We follow the same steps as in the case of the coherent state system of section 2.

First we start by writing the wavefunction as a linear combination of the energy eigenfunctions of the unperturbed system (Eq.~\eqref{lc}) and finding the eigenvalues (energy levels) and the eigenfunctions for every value of $\epsilon$, as well as their time evolution from $t=0$ up to $t=10000$.

In this case, in contrast with the coherent state, the wavefunction for $\epsilon=0$ contains only 3 low energy levels. Thus the introduction of a small $\epsilon\neq 0$ does not produce significant interaction of the high energy levels. For this reason it is sufficient to limit our truncation at $n_x+n_y=8$ {(the states with $n_x+n_y=8$ have probabilities of the order $10^{-9}$ for $\epsilon=0.01,$ $10^{-6}$ for $\epsilon=0.05$ and $10^{-5}$ for $\epsilon=0.09$)}.

\begin{figure}
\centering
\includegraphics[scale=0.18]{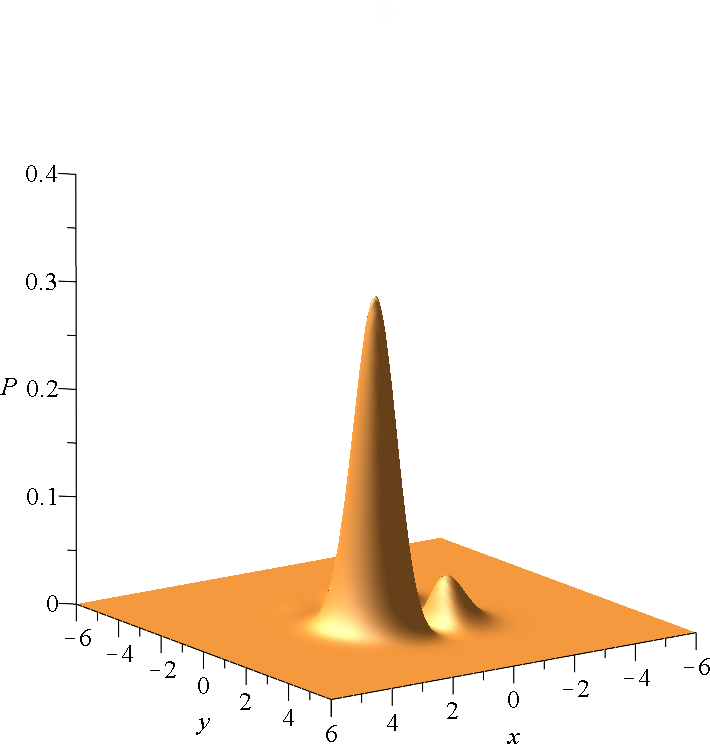}[a]
\includegraphics[scale=0.18]{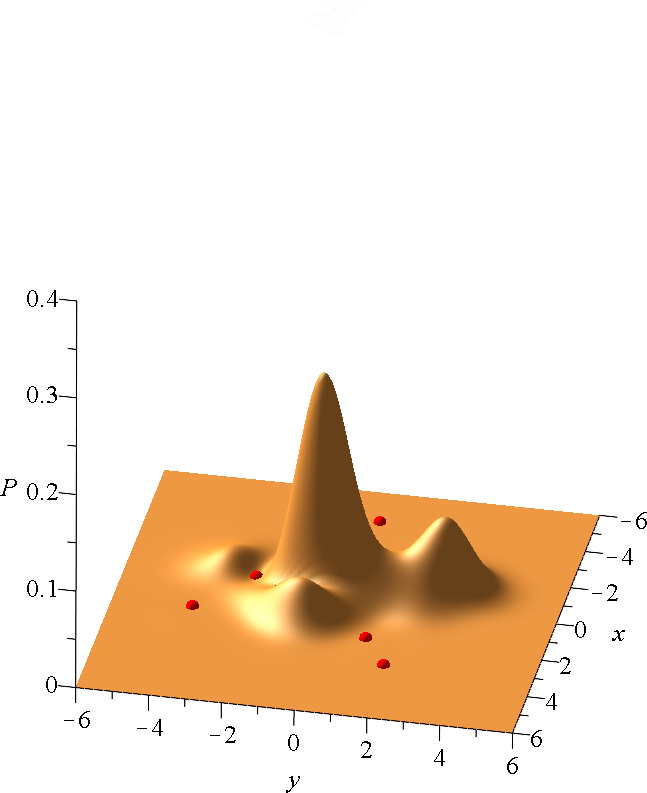}[b]
\caption{The probability density of the wavefunction \eqref{wfsn} at: a) $t=0$ and b) at $t=24$. }\label{psitetnod}
\end{figure}

The probability density follows Born's rule, i.e. $P=|\Psi|^2$ (Fig.~\ref{psitetnod}a,b). This gives a large blob and a secondary blob for $t=0$. As time increases these blobs move around the configuration space and they get deformed as in the cases of Figs.~\ref{snaps} (Fig.~\ref{psitetnod}b).

The values of the energy levels undergo irregular oscillations, as in Fig.~\ref{timeseries}. We find their average values $E_{av}$ by use of Eq.~(\ref{eaveq}) (but with 45 energy levels instead of 91) up to the time $t$ as in Fig.~\ref{eav}. The values of $E_{av}$ converge to values $\bar{E}$ as time increases up to the limiting value $t=10000$. These average values for every given value of $\epsilon$ are given in Fig.~\ref{ensingle} up to $\epsilon=0.1$.

\begin{figure}[H]
\centering
\includegraphics[width=0.48\textwidth]{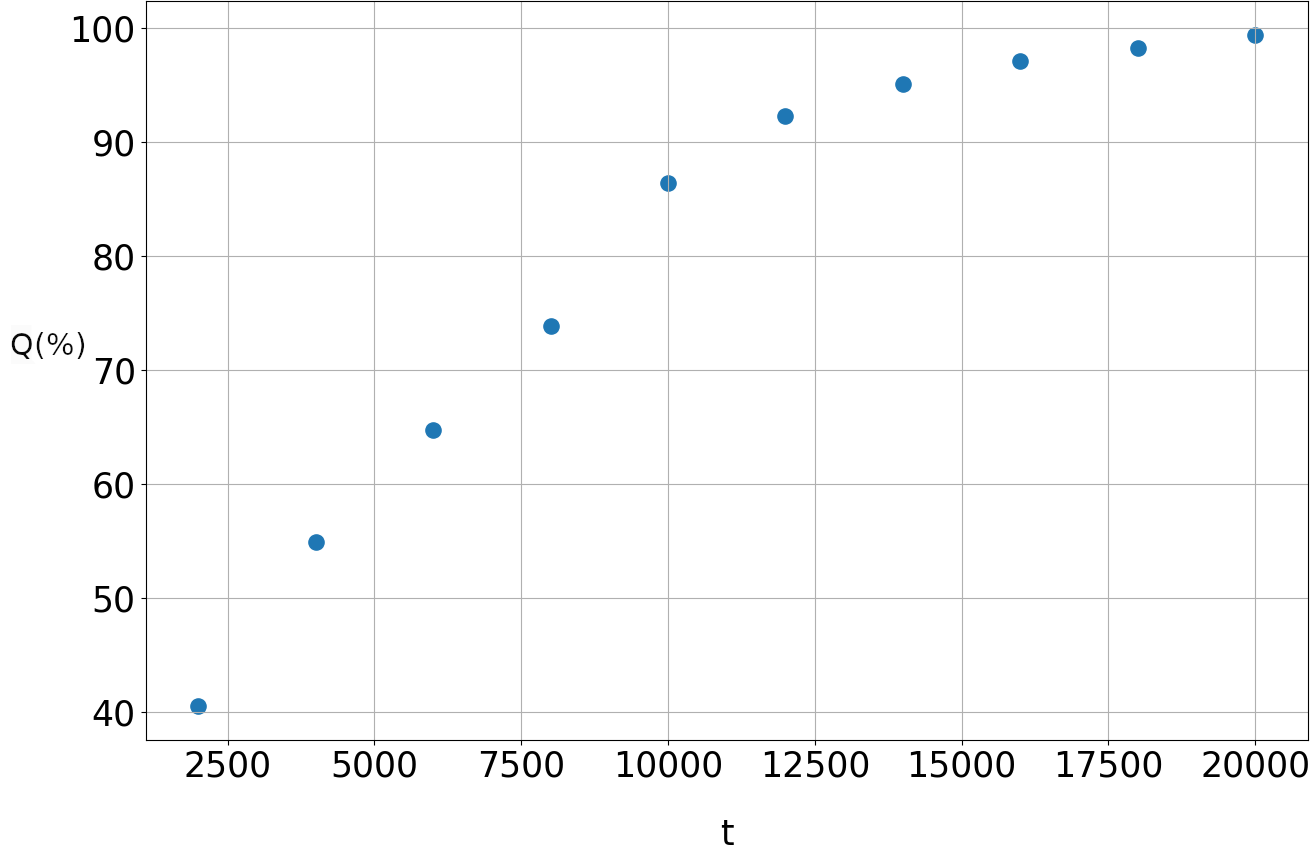}[a]
\includegraphics[width=0.48\textwidth]{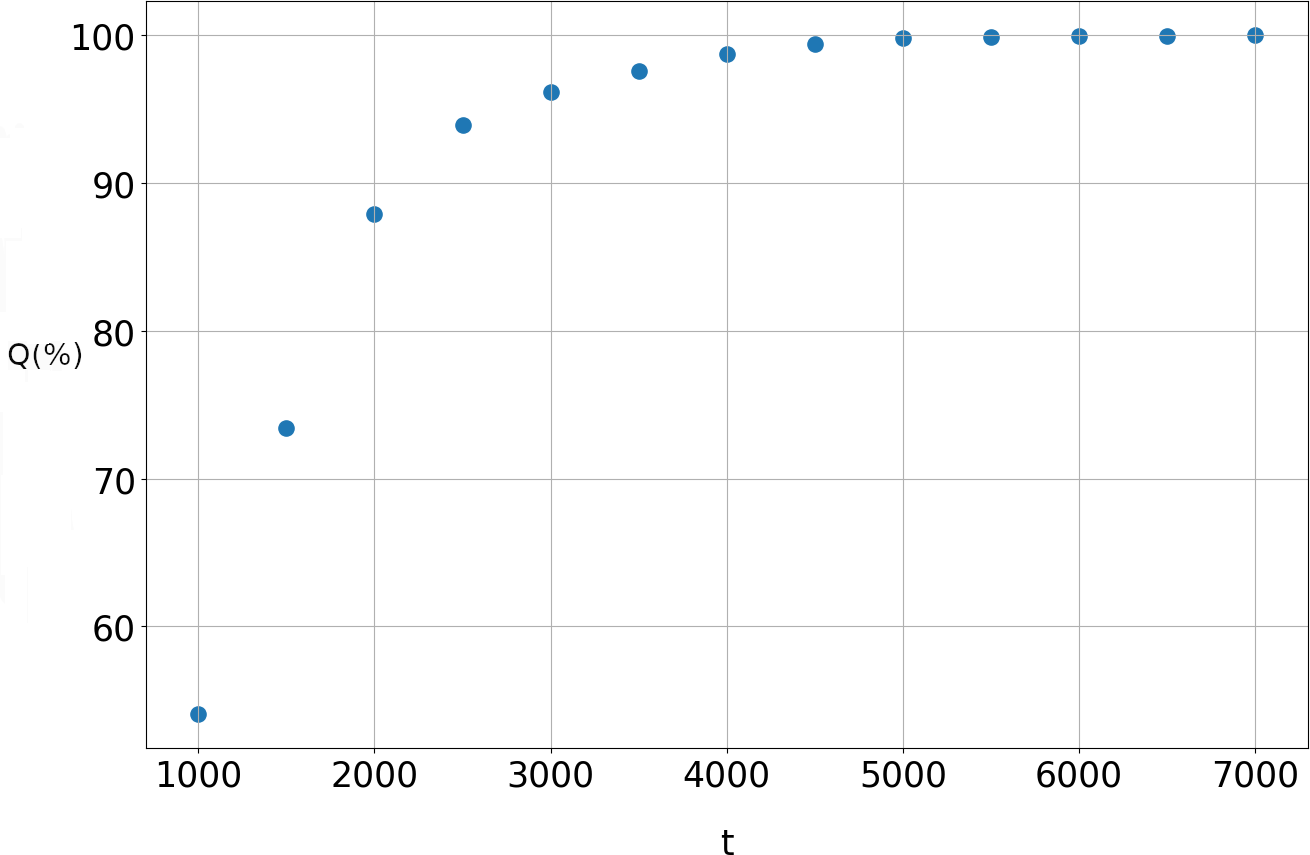}[b]
\caption{The proportion $Q$ of the Bohmian trajectories found to be chaotic in a realization of Born's rule distribution made out of 5000 particles as a function of time for: a) $\epsilon=0.05$ and  b) $\epsilon=0.09$.}\label{chaotic}
\end{figure}

We see that the values of $\langle E_{av}\rangle$ increase as $\epsilon$ increases but all these values are much smaller than the corresponding values of $\langle E_{av}\rangle$ of the coherent state case. E.g. for $\epsilon=0.09$ we have $\langle E_{av}\rangle=1.62$ while in the coherent state case we had $\langle E_{av}\rangle=3.29$.

%

For $\epsilon\neq 0$ there appear many nodal points besides the original one, which are at infinity for $t=0$, but later they enter the central region of space along with their associated X-points that deviate the approaching particles in a chaotic way.

The situation is similar with that of the coherent state but with a great difference in the time scales. Namely in the present case the times needed to establish the chaotic behaviour of the trajectories is much longer than in the coherent state. As an example, in order to reach approximately $100\%$ chaotic trajectories for $\epsilon=0.05$ we need a time $t\simeq 20000$ (Fig.~\ref{chaotic}a), while in the coherent state the corresponding time was very small ($t=200$). In the case $\epsilon=0.09$ this time is $t\simeq 4500$ (Fig.~\ref{chaotic}b) which is also very long compared with the corresponding time in the case $\epsilon=0.10$ of the coherent state ($t=35$). 

\begin{figure}[H]
\centering
\includegraphics[width=0.4\textwidth]{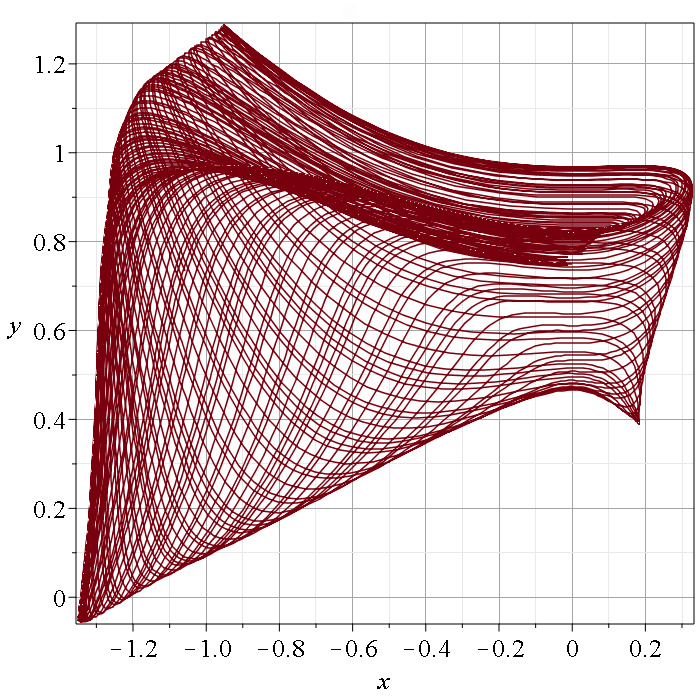}[a]
\includegraphics[width=0.4\textwidth]{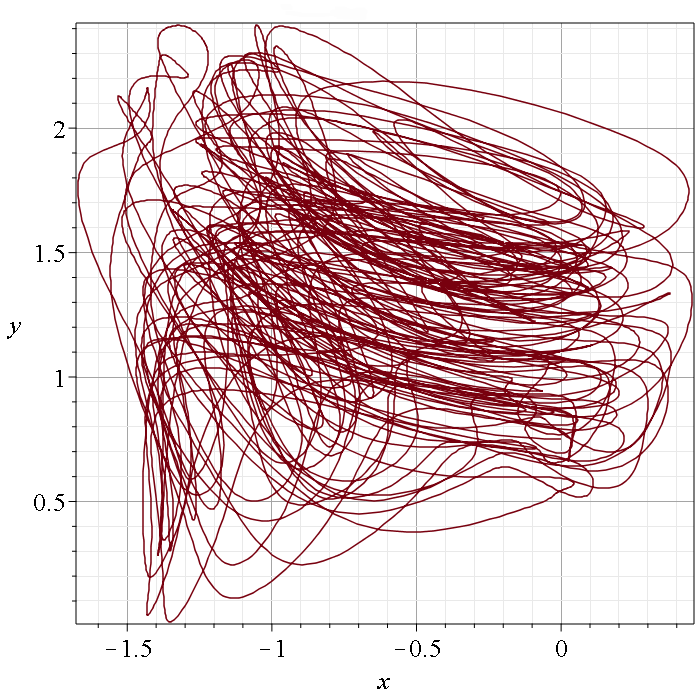}[b]
\caption{A Bohmian trajectory of the initial condition $x(0)=0, y(0)=0.75$ for: a) $\epsilon=0$ and b) $\epsilon=0.05$ up to $t=500$.}\label{trsingle}
\end{figure}

An example of the change of an ordered trajectory for $\epsilon=0$ to a chaotic trajectory for $\epsilon=0.05$ is shown in Figs.~\ref{trsingle}ab. In Fig.~\ref{trsingle}a the trajectory is a deformed Lissajous figure, while a trajectory with the same initial conditions for the same time ($t=500$) and $\epsilon=0.05$ is definitely chaotic (Fig.~\ref{trsingle}b).

We conclude that all the quantum trajectories for $\epsilon\neq 0$ are chaotic but the establishment of their chaotic form requires a long time, which is longer than the corresponding times of the coherent state.
Similar results are found for other wavefunctions that give some quantum chaos for $\epsilon=0$.

\section{Classical Hamiltonian chaos}

\begin{figure}[H]
\centering
\includegraphics[width=0.3\textwidth]{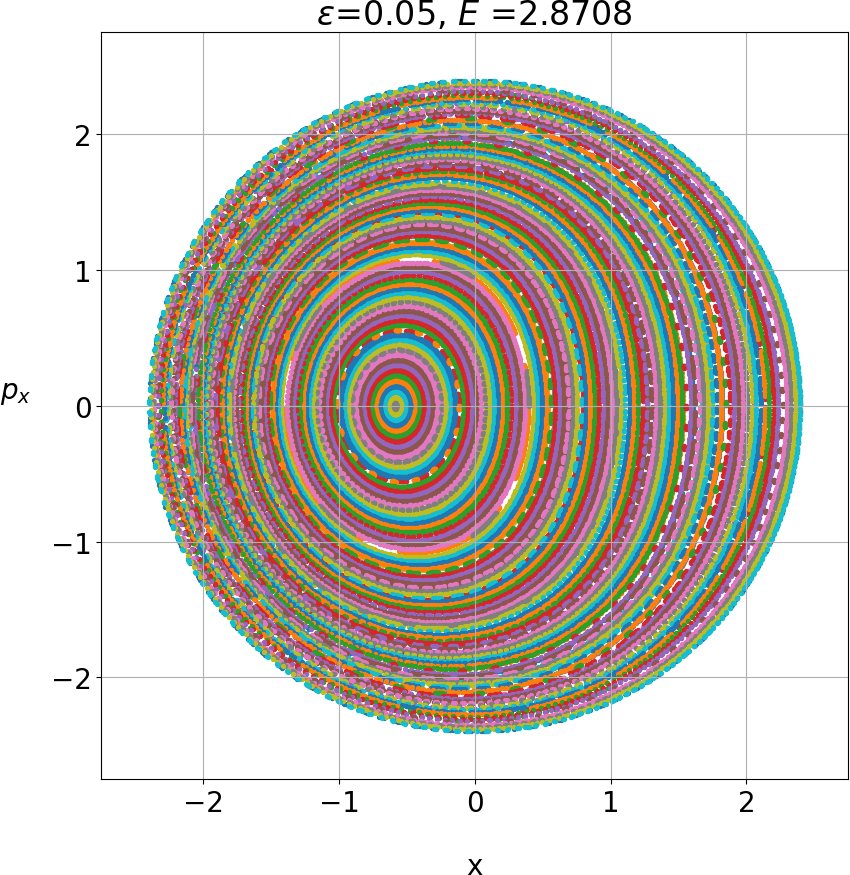}[a]
\includegraphics[width=0.3\textwidth]{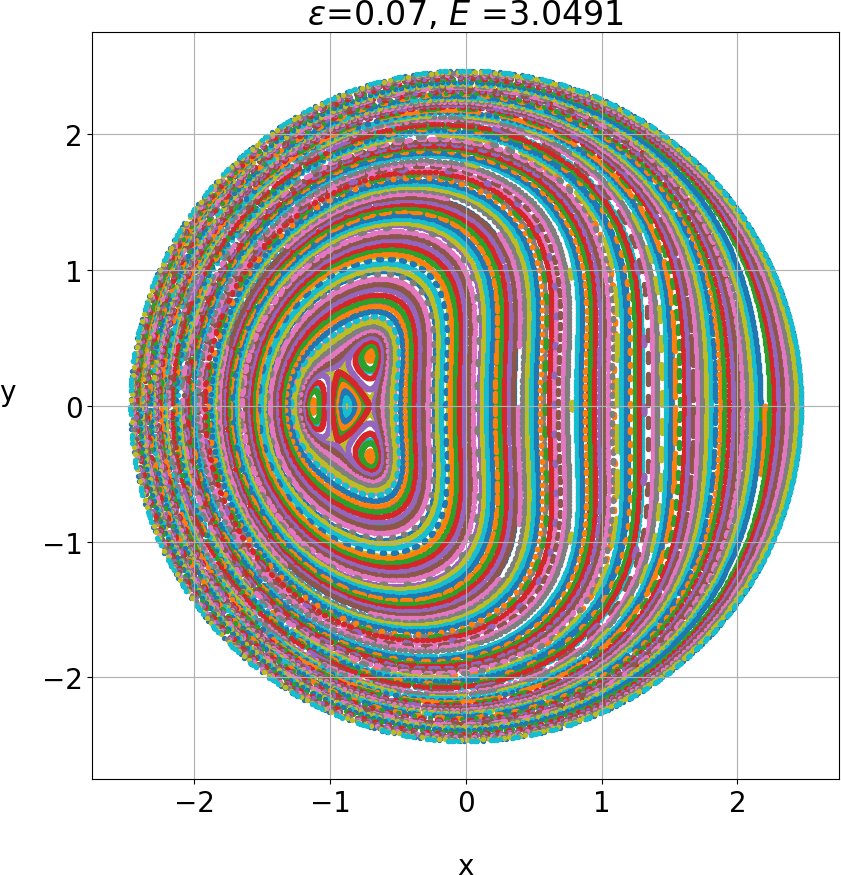}[b]\\
\includegraphics[width=0.33\textwidth]{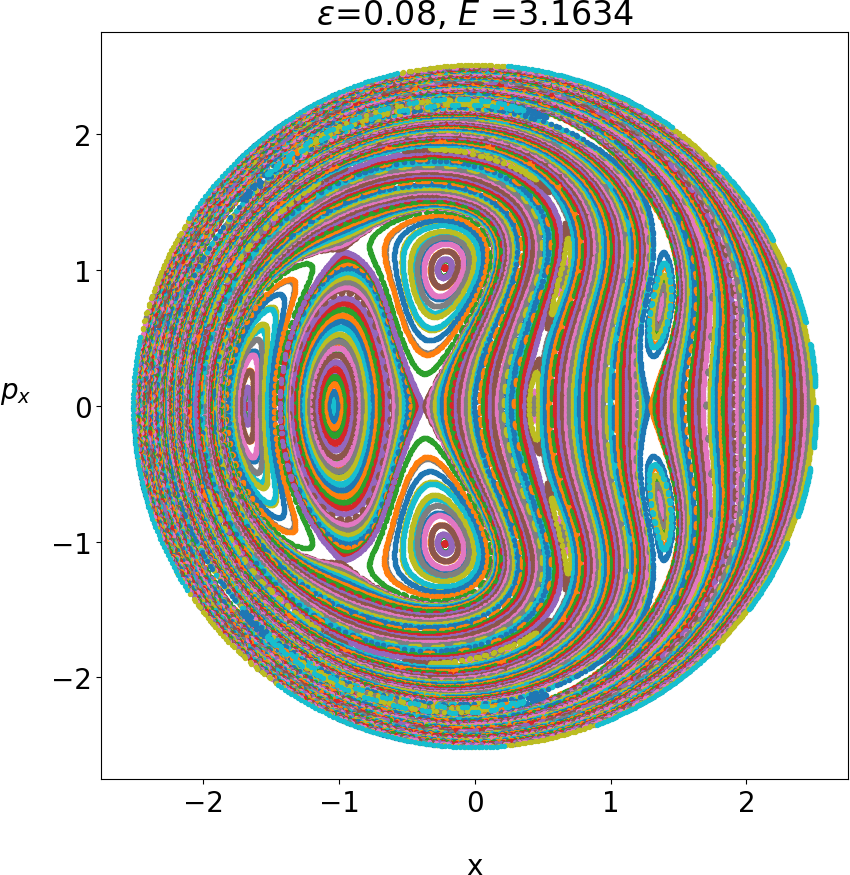}[c]
\includegraphics[width=0.3\textwidth]{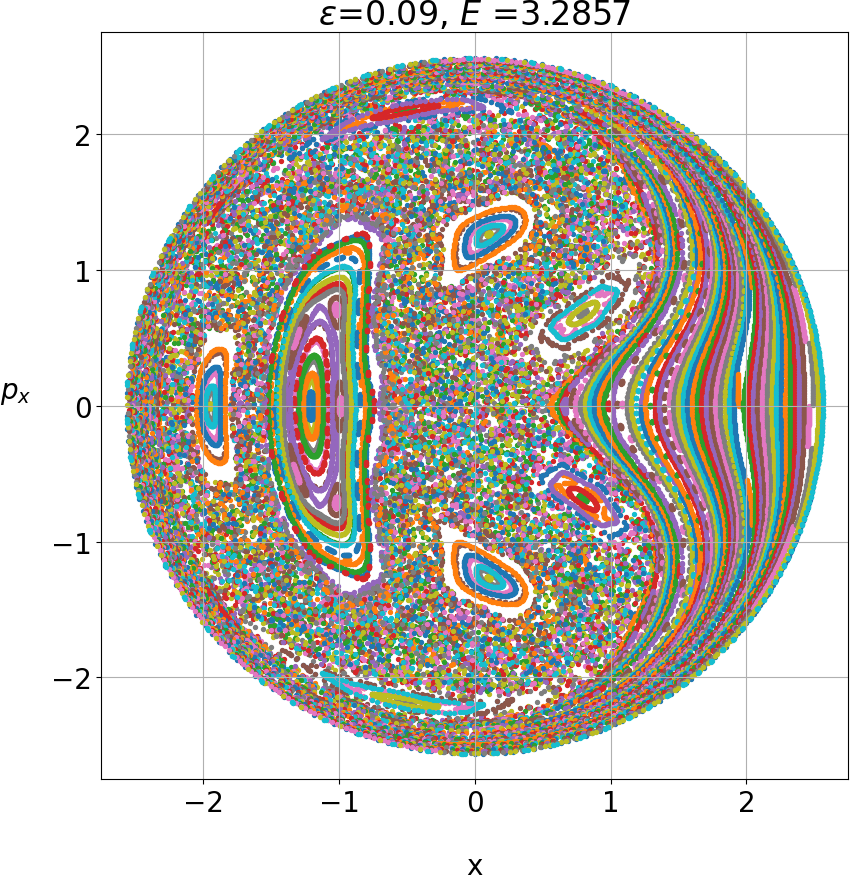}[d]\\
\includegraphics[width=0.3\textwidth]{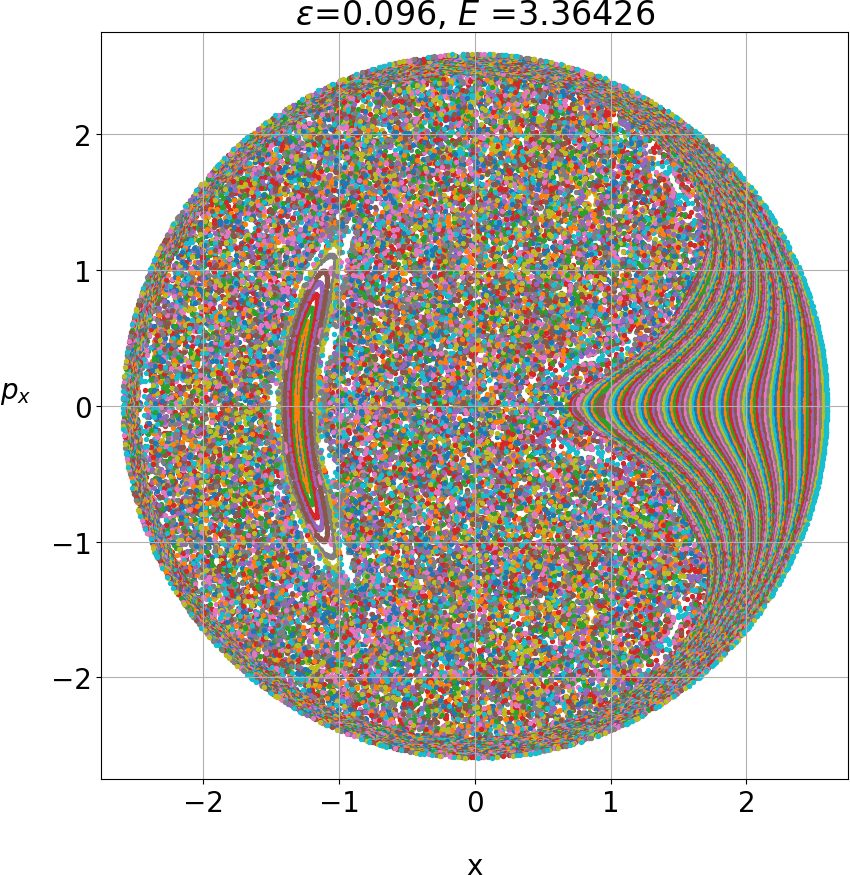}[e]
\includegraphics[width=0.3\textwidth]{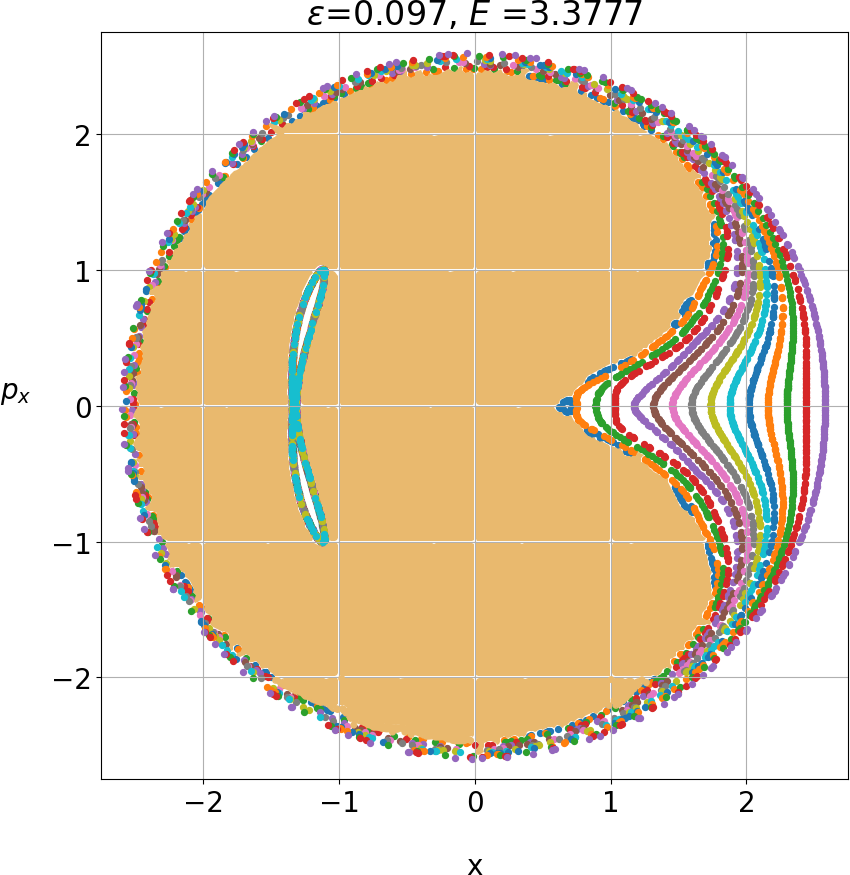}[f]
\caption{Poincar\'{e} surfaces of section ($y=0$) for 6 different values of $\epsilon$: a) $\epsilon=0.05$, b) $\epsilon=0.07$ c) $\epsilon=0.08$, d) $\epsilon=0.09$,  e) $\epsilon=0.096$ and f) $\epsilon=0.097$. The orange zone of f) refers to initial conditions that escape to infinity. The outermost invariant curve appears for $y=\dot{y}=0$ and it is a circle at the limits of a-e at $x=\pm\sqrt{2\langle E_{av}\rangle}$ namely at $x=\pm 2.4, \pm 2.47, \pm 2.52,\pm 2.56, \pm 2.59$ and $\pm 2.60$. It represents the stable periodic trajectory $y=0$. Trajectories starting at large $|x|$ for $y=0$ are approximately Lissajous figures and form invariant curves close to this circle. Such trajectories exist even when $\epsilon=0.097$ (f) and do not escape to infinity. The central periodic trajectory is at $x=-0.587, -0.882, -1.040,-1.210,-1.313$ and $x=-1.34$ respectively.  In the last case this periodic trajectory is unstable. In b), c), d) there is a stable triple unstable trajectory at $x=-1.129, 1.681$ and $x=-1.947$ correspondingly. Finally in b), c), d) and e) we find a triple unstable trajectory at $x=-0.709, -0.381, 0.064,$ and $x=0.105$ correspondingly.}\label{psos}
\end{figure}

The corresponding classical system is described by the Hamiltonian
\begin{equation}
H=\frac{1}{2}(\dot{x}^2+\dot{y}^2+\omega_x^2x^2+\omega_y^2y^2)+\epsilon xy^2=\langle E_{av}\rangle
\end{equation}
for the mean average values of $\langle E_{av}\rangle$ found in the quantum cases for every value of $\epsilon$.
Besides the integrals of the energy and of the angular momentum one can find a formal (i.e. approximate) integral of motion as series with respect to the power of the perturbation parameter $\epsilon$. This `third integral' which has been used extensively in Galactic Dynamics \cite{Contopoulos200210} takes, in the present case, the form:
\begin{equation}\label{phin}
\Phi=\frac{1}{2}(\dot{x}^2+\omega_x^2x^2)+\frac{\epsilon}{4\omega_y^2-\omega_x^2}[(2\omega_y^2-\omega_x^2)xy^2+2x\dot{y}^2-2y\dot{x}\dot{y}]+...=C
\end{equation}
On the surface of section $y=0$ this integral gives
\begin{equation}\label{phif}
\Phi=\frac{1}{2}(\dot{x}^2+\omega_x^2x^2)+\frac{2\epsilon x\dot{y}^2}{4\omega_y^2-\omega_x^2}+...=C,
\end{equation}
where $\dot{y}^2=2\langle E_{av}\rangle-\omega_x^2x^2-\dot{x}^2$. 
The curves \eqref{phin} are invariant curves. Such curves are given in Figs.~\ref{psos}abcde for $\omega_x=1,\omega_y=\sqrt{2}/2$, and various values of $\epsilon$ along with the corresponding $\langle E_{av}\rangle$ (Fig.~\ref{eav}b), while $C$ takes a constant value for every invariant curve.  

\begin{figure}[H]
\centering
\includegraphics[scale=0.2]{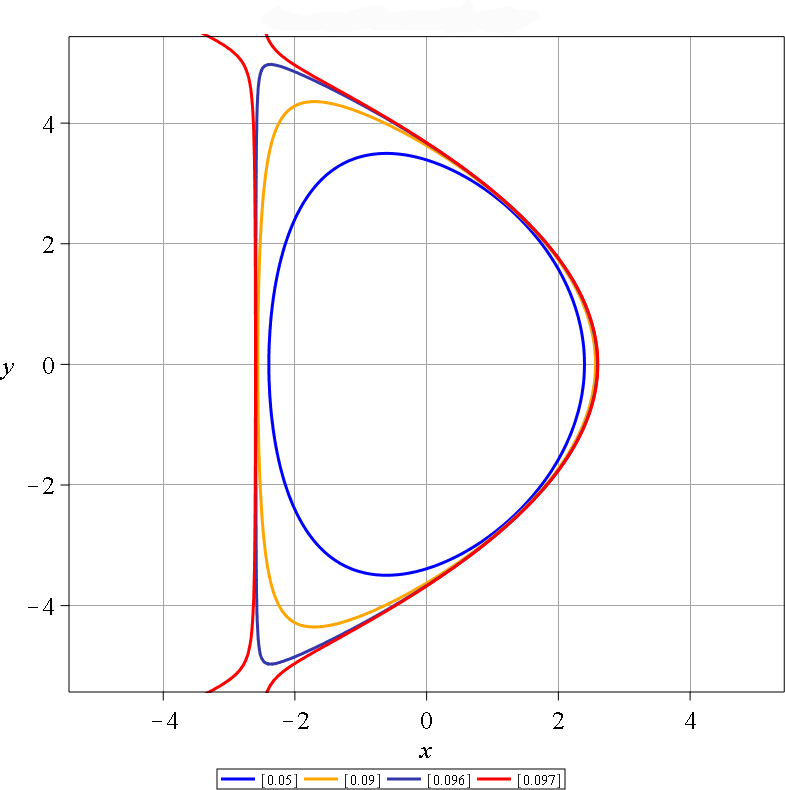}
\caption{The curves of zero velocity for $\epsilon=0.05$ (central blue curve), $\epsilon=0.09$ (orange curve), $\epsilon=0.096$ (dark blue curve) and $\epsilon=0.097$ (red open curve).  The escape perturbation is between $\epsilon=0.096$ and $\epsilon=0.097$.}\label{CZVs}
\end{figure}

The corresponding classical trajectories are surrounded by a `curve of zero velocity' (CZV):
\begin{equation}
\frac{1}{2}(\omega_x^2x^2+\omega_y^2y^2) +\epsilon xy^2=\langle E_{av}\rangle,
\end{equation}
which in the present  case is approximately an oval 
\begin{equation}
y^2=\frac{2\langle E_{av}\rangle-\omega_x^2x^2}{\omega_y^2+2\epsilon x}.
\end{equation}
Some  CZVs are given in Fig.~\ref{CZVs}. We see that for $\epsilon\geq 0.097$ the CZV is open and most trajectories escape to infinity. 

The perturbation which starts to give open CZVs corresponds to an escape energy  $E(\epsilon_{esc})$. We note that the Bohmian trajectories are still bounded even when $\epsilon>\epsilon_{esc}$, since the quantum system is always bounded.
By taking now only the first order terms of Eq.~\eqref{phif} in $\epsilon$ we have 
\begin{align}
\Phi=\frac{1}{2}(x^2+\dot{x}^2)(1-4\epsilon x)+4\epsilon \langle E_{av}\rangle x=C.
\end{align}
In particular, when 
\begin{equation}
\frac{\partial \Phi}{\partial \dot{x}}=\frac{\partial \Phi}{\partial {x}}=0,
\end{equation}
the invariant curve shrinks to a point which represents the central periodic trajectory (stable for $\epsilon\leq 0.085$). This point has $\dot{x}=0$ and
 in the lowest approximation
\begin{equation}
x=-4\langle E_{av}\rangle\epsilon.
\end{equation}

\begin{figure}[H]
\centering
\includegraphics[width=0.45\textwidth]{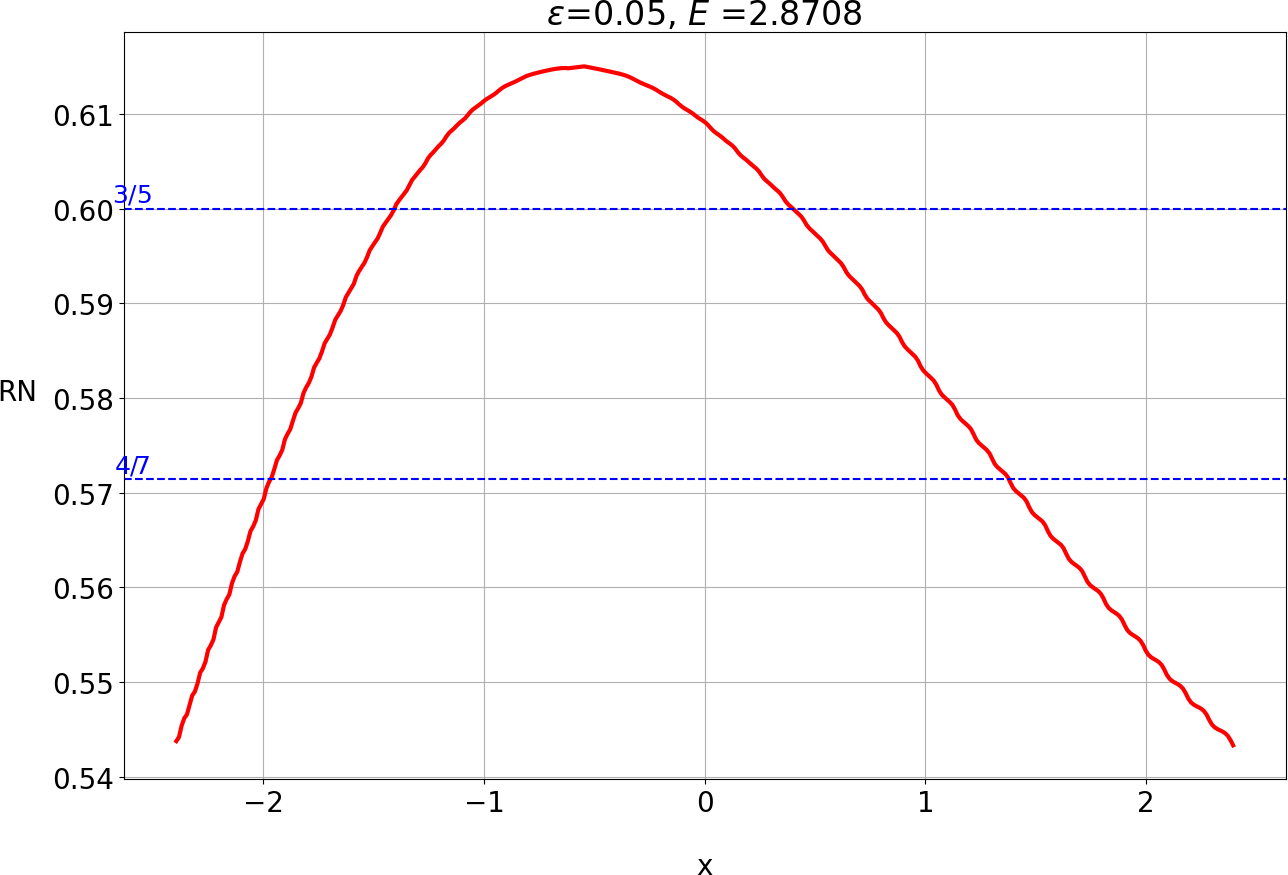}[a]
\includegraphics[width=0.45\textwidth]{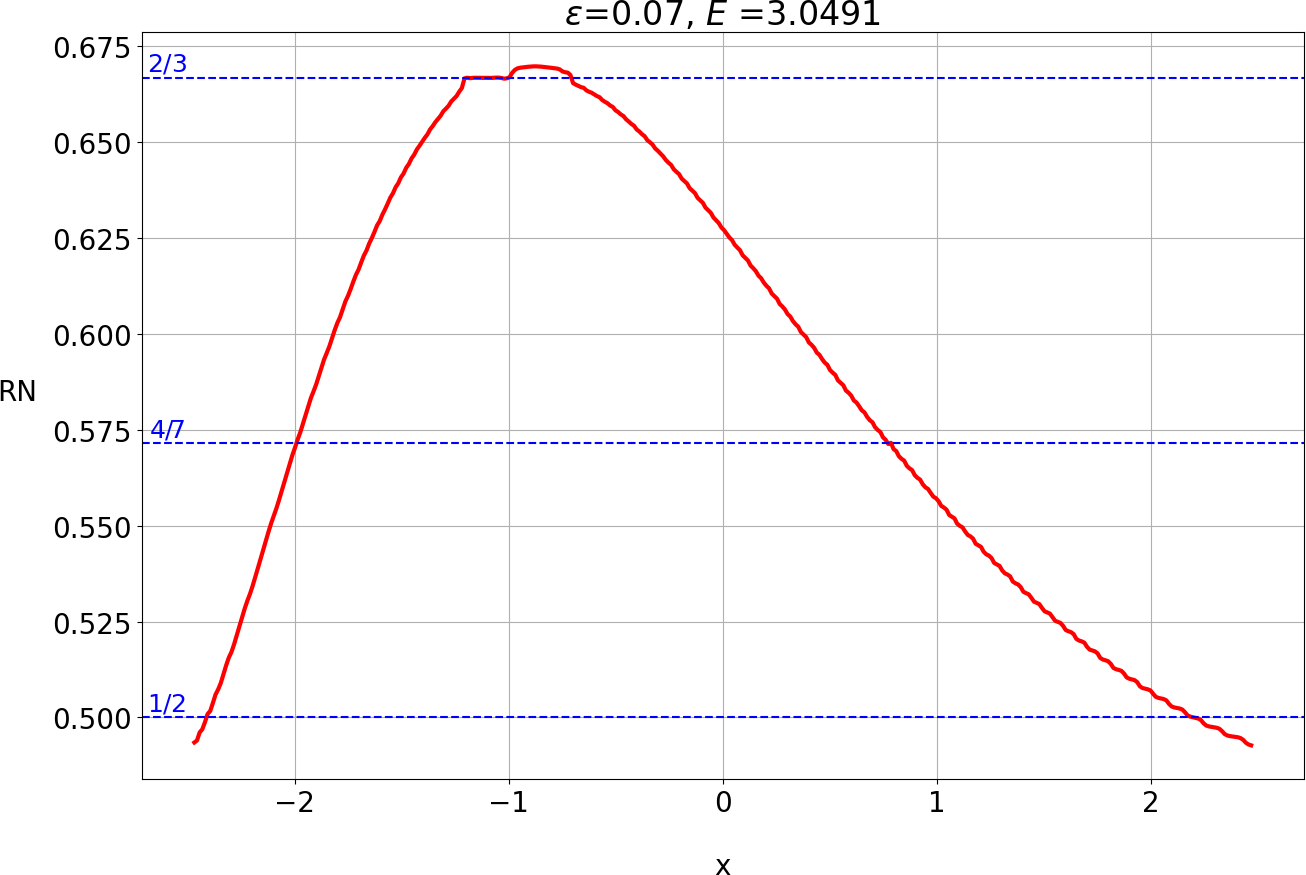}[b]
\includegraphics[width=0.45\textwidth]{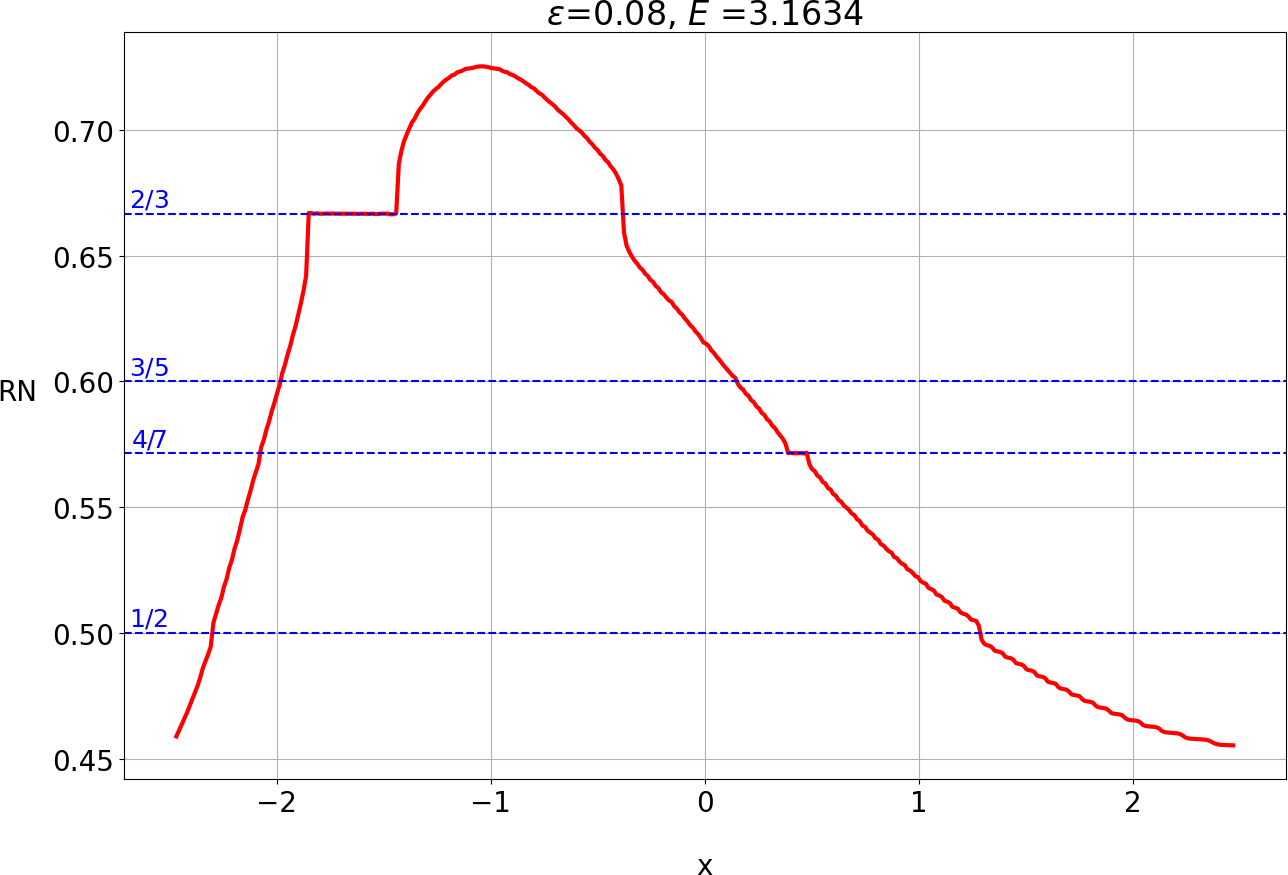}[c]
\includegraphics[width=0.45\textwidth]{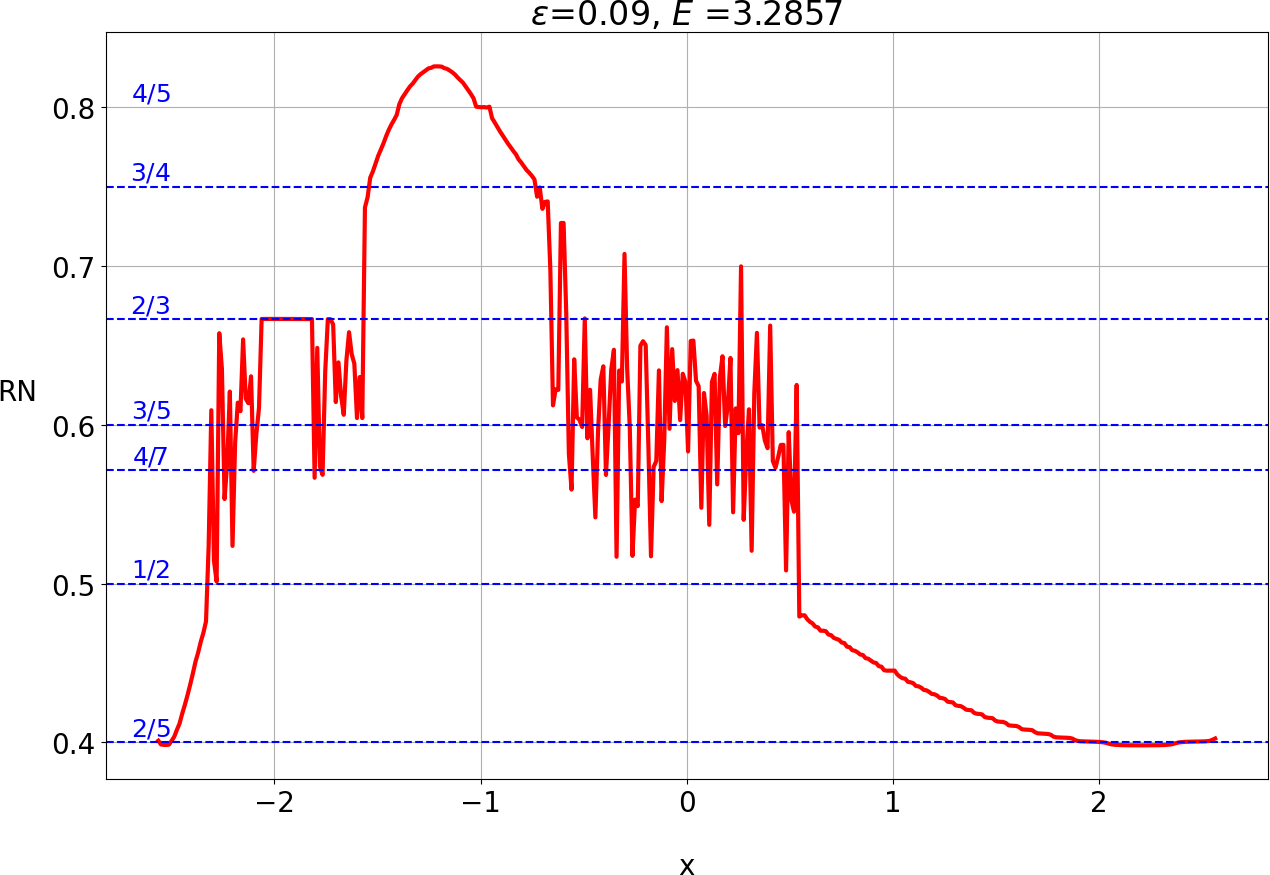}[d]
\includegraphics[width=0.55\textwidth]{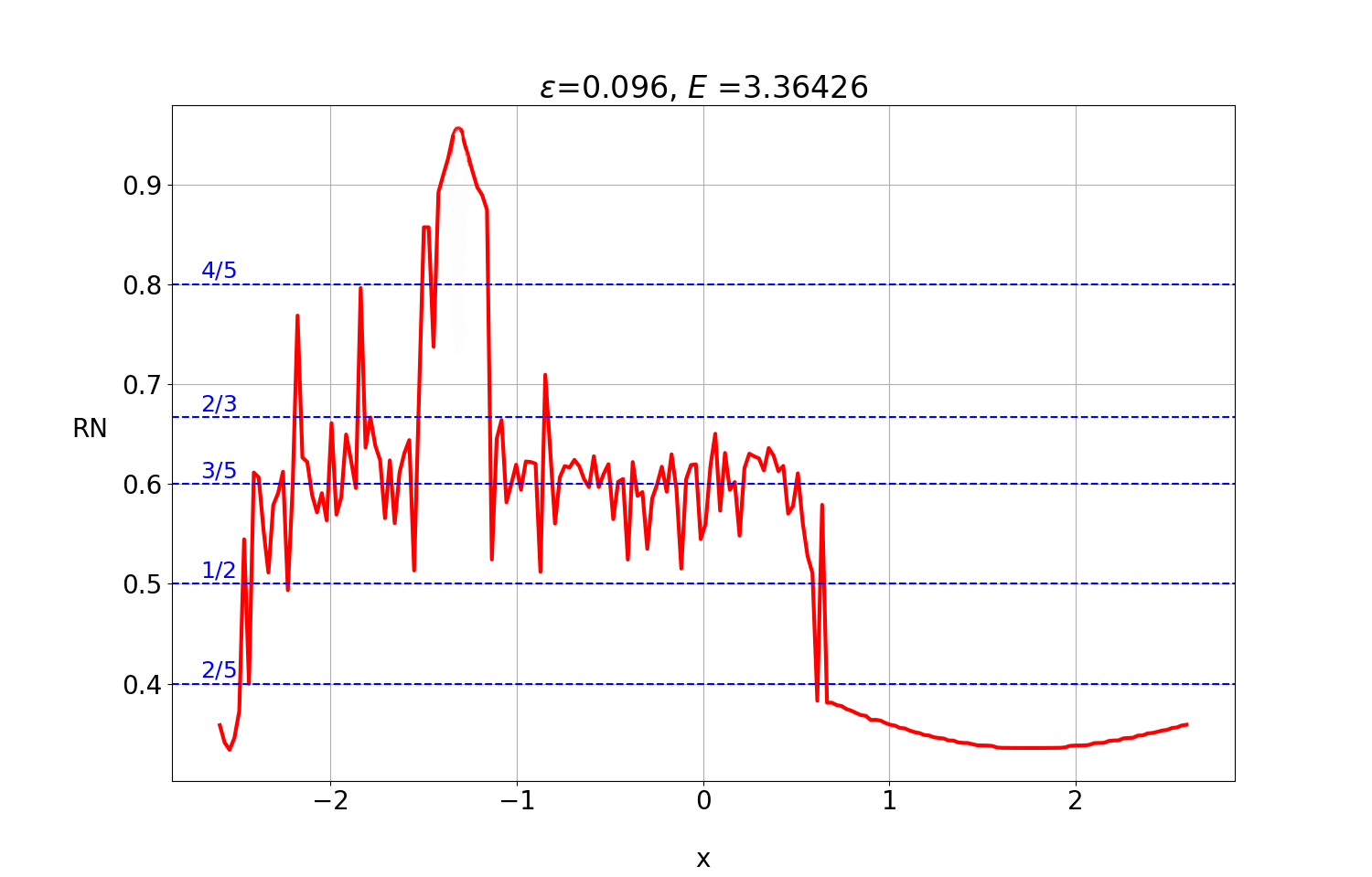}[e]
\caption{The rotation number $RN$ for various values of $\epsilon$: a) $\epsilon=0.05$,  b) $\epsilon=0.07$, c) $\epsilon=0.08$ and d) $\epsilon=0.09$. The maximum values are  $RN=0.62, 0.67, 0.73, 0.83$ and $0.95$ respectively. The minimum values are near $RN=0.54, 0.49$ and $0.46$ in a,b,c respectively, while they are a little before the limits of (d) at about $RN=0.40$ ($\epsilon=0.09$) and $RN=0.34$ for $\epsilon=0.096$ (e).}\label{RN}
\end{figure}

Besides the central periodic trajectory there are several stable and unstable periodic trajectories that are found when the `rotation number' ($RN$) of the invariant curves is  rational (the rotation number is the average value of the  azimuth angles $\Delta\phi$ of the successive points of an invariant curve as seen from the central periodic trajectory point) \cite{Contopoulos200210}. The rotation number as a function of $x$ is given in Figs.~\ref{RN}abcde.

\begin{figure}[H]
\centering
\includegraphics[width=0.4\textwidth]{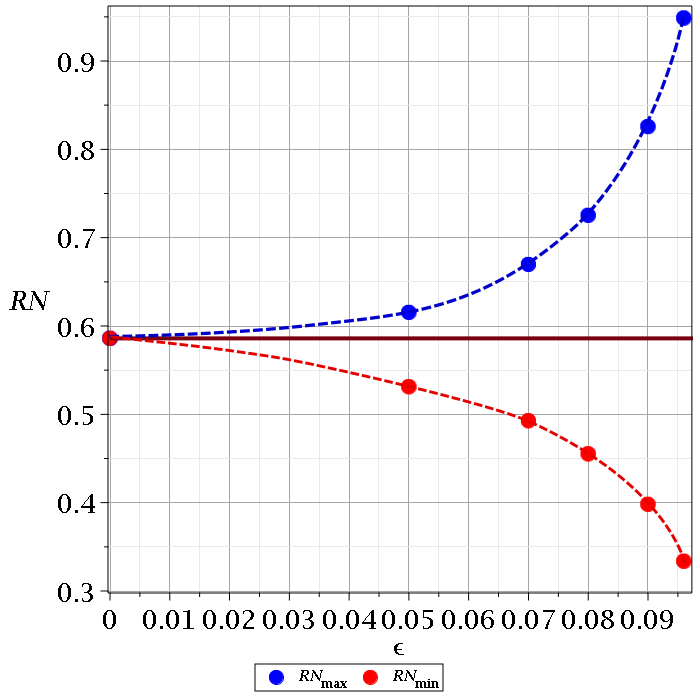}
\caption{The extrema of the rotation number $RN$ for various values of $\epsilon$.}\label{RNext}
\end{figure}

The basic rotation number is $RN=2-\omega_x/\omega_y=0.586$ and it remains constant for all the trajectories if $\epsilon=0$.  As $\epsilon$ increases the maximum rotation number increases (Fig.~\ref{RNext}).   The outermost curves of Figs.~\ref{psos}abc correspond to the minima of the  $RN$ in Figs.~\ref{RN}abc.  The maxima and minima of $RN$ are given in Fig.~\ref{RNext} as functions of the perturbation parameter $\epsilon$.

For increasing $\epsilon$  the rotation number goes through various rational values. The most important resonant number is $RN=2/3$. The rotation curve ($RN$ as a function of $x$) has a plateau for $RN=2/3$ on the left of the center and an abrupt change on the right of the center in Figs.~\ref{RN}b,c,d. The plateaus are around a triple stable periodic trajectory.  For $\epsilon=0.096$ this stable triple periodic trajectory has become unstable. In fact, this periodic trajectory is a bifurcation of the central periodic trajectory and appears for the first time when $\epsilon$ goes above $0.069$. Between the 3 islands around the stable periodic trajectory there are 3 points representing a triple unstable periodic trajectory (Figs.~\ref{psos}abcd). Near the 3 unstable points there are small chaotic regions in Figs.~\ref{psos}bc, ($\epsilon=0.05, 0.07$) which become a large chaotic domain for $\epsilon=0.09$ and $\epsilon=0.096$ (Fig.~\ref{psos}de).

{In Figs.~\ref{psos}c,d,e there are further islands around stable periodic
trajectories. Moreover, there exist unstable periodic trajectories that
contribute to chaos. These periodic trajectories correspond to various
resonances in Figs.~\ref{RN}c,d,e (e.g. resonances 1/2, 3/4, 4/5, 4/7).}

\begin{figure}[H]
\centering
\includegraphics[scale=0.2]{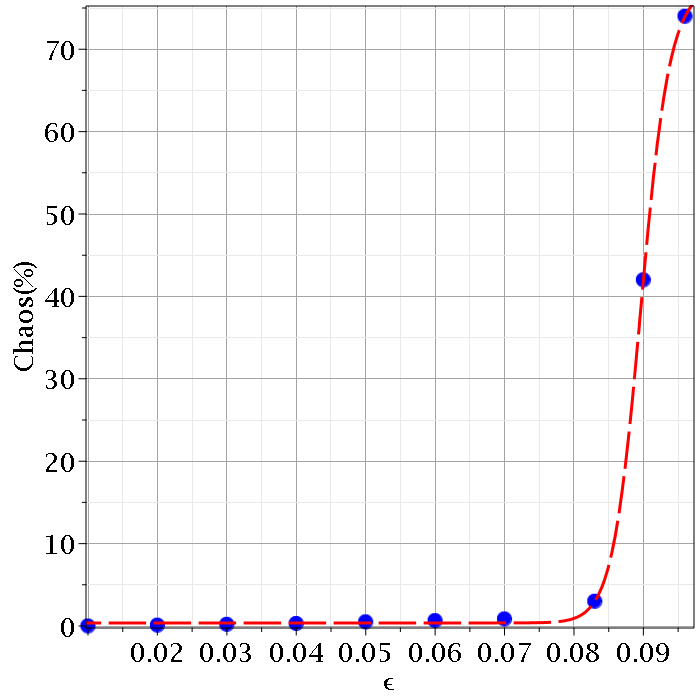}
\caption{The area covered by chaotic trajectories as a function of the perturbation $\epsilon$.}\label{chaos_area}
\end{figure}

The amount of chaos can be estimated by the percentage of the chaotic areas in Figs.~\ref{psos}abcde and in intermediate figures (with respect to the total area). This percentage is very small for $\epsilon\leq 0.082$ and increases abruptly for $\epsilon>0.083$ (Fig.~\ref{chaos_area}). At $\epsilon=0.083$ it is $ch=3\%$, at $\epsilon=0.09$ it is $ch=43\%$ and at $\epsilon=0.096$ it is $74\%$. For $\epsilon\geq 0.097$ most trajectories escape to infinity and only a few ordered trajectories remain around two islands above and below the central periodic trajectory and close to the boundary circle (Fig.~\ref{psos}f) which corresponds to the stable periodic trajectory $y=0$.

%

The increase of chaos is mainly due to the overlapping of resonances. E.g. at about $\epsilon=0.085$ starts an overlapping of the resonances 2/3 and 3/5, while at $\epsilon=0.09$ we have an overlapping of the resonances $3/4, 2/3, 3/5, 4/7$ and $1/2$.

When the rotation curve has an abrupt change near an unstable periodic trajectory with $RN=n/m$ there is an overlapping of resonances with $RN$ close to $n/m$. The overlapping is larger as $\epsilon$ increases and affects larger intervals of values of $RN$ around $n/m$. When two distant simple ratios $RN$ overlap (like $RN=2/3$ and $1/2$ in Figs.~\ref{psos}d, \ref{RN}d) the chaotic domain is large.

The phenomenon of resonance overlap was introduced in 1966 by Rosenbluth et al. \cite{rosenbluth1966destruction} and Contopoulos \cite{contopoulos1967resonance} and has been studied in  detail by Chirikov \cite{chirikov1979universal}. Resonance overlap appears when the unstable asymptotic curves of one resonant periodic trajectory intersect the stable asymptotic curves of another resonant periodic trajectory. Then a set of periodic trajectories close to the first periodic trajectory come close to the second periodic trajectory and these are chaotic (for details see \cite{Contopoulos200210}). Here we see various cases of resonance overlap, with emphasis on those in the cases of Figs.~\ref{psos}d,e.

In the case of a single nodal point (section 3) the values of the energies $\langle E_{av}\rangle$ are given in Fig.~\ref{ensingle} and they are smaller than in the case of the coherent state (Fig.~\ref{eav}b). This change affects the values of $\langle E_{av}\rangle(\epsilon)$ to be used in providing surfaces of section like those of Fig.~\ref{psos} and curves of zero velocity like Fig.~\ref{CZVs}. In particular the average escape perturbation in this case is $\epsilon_{esc}=0.132$, larger than the value given above ($\epsilon_{esc}=0.083$) for the coherent case. As a consequence the classical chaos for any given $\epsilon$ is smaller than that given above for the coherent state and the transition to large chaos appears for larger $\epsilon$. Usually in classical mechanics we fix either the value of energy and find a fixed escape perturbation or the value of $\epsilon$ and find a fixed escape energy \cite{Contopoulos200210}. Here however, we have to change both $\epsilon$ and $E$ in order to compare the classical with the quantum trajectories.

%

\section{Discussion and Conclusions}

The purpose of this paper is to compare the classical and the Bohmian quantum chaos.
We use a simple classical potential $V=\frac{1}{2}(\omega_x^2x^2+\omega_yy^2)+\epsilon xy^2$ with non-commensurable frequencies $\omega_x=1, \omega_y=\sqrt{2}/{2}$ and values of $\epsilon$ from $\epsilon=0$ up to the escape perturbation. For these values of $\epsilon$ the phase space of the classical system is of finite size. For larger values most particles escape to infinity.
The corresponding quantum systems are various solutions of the Schr\"{o}dinger equation for every value of $\epsilon$.  The quantum solution exists even for larger values of $\epsilon$ than that of the classical escape perturbation.

We worked with a wavefunction which for $\epsilon=0$ produces only ordered trajectories which are close to those of the corresponding classical case and with a wavefunction where order and chaos coexist for $\epsilon=0$. In particular the first wavefunction is a product of two coherent states along the coordinates $x$ and $y$ with Poissonian probability distributions for the various energy levels. The distribution of the Bohmian particles is assumed to follow  Born's rule $P=|\Psi|^2$  and forms a blob of the values of the probability density $P(x,y)$ (Fig.~\ref{poisson}b). If $\epsilon=0$ the blob moves around in time but its form is preserved. On the other hand, in the second wavefunction there is a proportion of chaotic Bohmian  trajectories already when $\epsilon=0$.

In both cases we made a numerical diagonilization of the Hamiltonian of the  perturbed systen and then we wrote its  wavefunction as a linear combination of the energy eigenfunctions of the unperturbed problem truncated at   a large enough number of energy levels. We then found the contribution of the various perturbed energy levels in the time evolution of the wavefunction. 


Our main result is that in both cases the Bohmian trajectories are in general chaotic for $\epsilon>0$. In particular, in the first case the trajectories form  approximate Lissajous figures for some time (close to the Lissajous figures of the unperturbed system), {but later they approach the nodal points and their associated X-points and become chaotic}. By considering a realization of Born distribution at $t=0$ we found that the time needed for the proportion of chaotic trajectories to reach $100\%$ is  small for large $\epsilon$ but it increases considerably as $\epsilon$ decreases, tending to infinity when $\epsilon\to 0$. The same results were found in the second case as well but at much larger times for any given value of $\epsilon$.


%

%

As regards the classical case, it is well known that chaos is introduced when particles approach an unstable periodic trajectory. We found the distribution of the chaotic and the ordered trajectories on a Poincar\'{e} surface of section ($y=0$ with $\dot{y}>0$) for various values of $\epsilon$, and  for values of the total energy equal to that of the mean average energy $\langle E_{av}\rangle$  of the corresponding quantum system up to a large time.

The set of the ordered trajectories around the main stable periodic trajectory  is given by a formal (approximate) integral of motion. By use of this integral and of the energy integral we found the invariant curves on the plane $(x,\dot{x})$ for $y=0$. For particular values of $x$ and $\dot{x}$ these curves are reduced to  points which represent a central periodic trajectory.

Every invariant curve around the central point has a certain rotation number ($RN$) (average angle of the successive points of a trajectory at $y=0$, as seen from the central point). The rotation number is constant ($R_0=2-\frac{\omega_x}{\omega_y}=2-\sqrt{2}$ in the case $\epsilon=0$ but it varies between a minimum and a maximum value as a function of $x$ (and $\dot{x}=0$) for every $\epsilon\neq 0$. The range of values of RN increases as $\epsilon$ increases. At every rational value of RN there are, in general, a stable  and an unstable periodic trajectory.

The most important resonant trajectories are those  with $RN=2/3$ and  $1/2$. Near every unstable periodic trajectory there is some chaos.

We found that classical chaos is extremely small for weak interactions ($\epsilon\leq 0.082$) but increases abruptly beyond $\epsilon=0.083$. The increase of chaos is due to the overlapping of resonances. Overlapping occurs when the invariant KAM curves separating  two resonances are destroyed and the asymptotic curves from the periodic trajectories of one resonance intersect those of the other resonance (stable vs unstable asymptotic curves). Then the trajectories close to one resonant periodic trajectory may go close to the other resonant trajectory. The overlapping for relatively large $\epsilon$ affects also rather distant resonances like $2/3$ and $1/2$.

As a consequence in the classical case in general we have both ordered and chaotic trajectories (except if $\epsilon=0$) but chaos becomes important only beyond a critical value of the perturbation. On the other hand in the quantum cases chaos appears for all the values of $\epsilon\neq 0$. Even if for $\epsilon=0$ there is no Bohmian chaos (as in the case of the coherent state), for every value of $\epsilon\neq 0$ all the trajectories are chaotic.


We note that  Bohmian chaos occurs due to the approach of the particles to the region of a nodal point, i.e a point where $\Psi=\Psi_{Real}+i\Psi_{Imag}=0$. Close to every nodal point $N$ there is an unstable point $X$ in the frame of reference centred at $N$. Bohmian particles approaching $X$ are deflected along the two opposite directions of the unstable asymptotic curves of $X$, thus nearby trajectories can be separated considerably. (There are also unstable points $Y$ in the inertial frame of reference, that contribute to chaos but at a smaller degree).
This means that, although the Bohmian trajectories differ significantly from the classical trajectories, their chaotic behaviour is produced by a similar mechanism.

However, the distribution of the  unstable points is very different in the classical and in the quantum case.
Namely, in the classical case the unstable points are in general dense (except in the integrable cases) and large chaos appears when the asymptotic curves of various unstable trajectories overlap. In the quantum case the nodal points, the X-points and the Y-points  are well separated in general, although they may collide from time to time. Thus there is no resonance overlap, as in the classical case.



A comparison between the classical and the Bohmian  chaotic trajectories can be summarized as follows:
\begin{enumerate}
\item Between the classical chaos and the Bohmian quantum chaos there is one basic similarity. In both cases the trajectories become chaotic when they approach an unstable periodic point (unstable periodic  trajectory).
\item However between the two cases there are several important differences. Considering the intersections of the trajectories by a surface of section ($y=0$) we note that :
\begin{enumerate}
\item The periodic trajectories in the classical case are dense (this is well known \cite{Contopoulos200210}), while in the quantum case they are in general separated. The quantum unstable points that produce most of chaos are the X-points near the nodal points $N$ (in the frame of reference centred at the nodal points). A minor role is played by the unstable points of the inertial frame of reference, the Y-points.
\item In the classical case near every unstable point there are unstable points of higher multiplicity that overlap (i.e. their asymptotic curves cross each other (stable with unstable). When two low order resonances overlap there is a large degree of chaos. 

\item On the other hand, in quantum systems there is no resonance overlap and most trajectories for $\epsilon\neq 0$ are chaotic. There is only a difference of the time when the chaotic character of the trajectories is established. Namely, when $\epsilon$ is large this time is short, while for small $\epsilon$ chaos emerges at a larger time, that tends to infinity when $\epsilon\to 0$. For a fixed time $t=t_c$ we have a corresponding critical perturbation $\epsilon_c$ such that for $\epsilon<\epsilon_c$ the trajectories look ordered (perturbed Lissajous figures) and for $\epsilon>\epsilon_c$ most trajectories are established as chaotic. This $\epsilon_c$ is similar to the transition perturbation of the classical system. However, this critical perturbation differs from the classical critical perturbation because it depends on the assumed time $t_c$. But if we consider a large time then we can distinguish between apparently ordered and apparently chaotic cases as in the classical case, although for even larger times we know that the Bohmian trajectories become in general chaotic.

\end{enumerate}
\item For every value of $\epsilon\neq 0$ there is a corresponding value of the average energy $\langle E_{av}\rangle$ of the eigenstates of a quantum system and there is no escape perturbation $\epsilon$. However in the corresponding classical system there is an escape perturbation $\epsilon_{esc}$ and for larger $\epsilon$ most classical trajectories escape to infinity, while  in the quantum system there are no escaping trajectories.
\end{enumerate}

In the present paper we worked with a simple Hamiltonian containing a non-linear coupling between 2 harmonic oscillators.

{We found that in the long run practically all Bohmian trajectories
exhibit chaotic behaviour when $\epsilon \neq 0$, while in the classical case we
have both ordered and chaotic trajectories for every value of $\epsilon$ (we
have large chaos only above a critical value of $\epsilon$). In our previous
papers \cite{tzemos2020ergodicity,tzemos2020chaos,tzemos2021role} we showed that the Bohmian chaotic trajectories
are in general ergodic, i.e. they have a common long time distribution
of points in the configuration space. This implies that, in the present
case, Born’s rule is in general accessible to all initial distributions of
Bohmian particles. But further work needs to be done to check the
generality of our present results in cases with different potentials.}

\section*{Appendix A: The diagonalization process}

In the introduction we pointed out that the choice of the initial quantum state affects the number of the energy eigenstates $|\Phi_m\rangle$ (of the unperturbed system) that one should take into account in the state $|\Psi(t)\rangle$ of the perturbed system. In the case of the coherent state we worked with all doublets of $n_x, n_y$ which sum up to 12, while in the second case we worked with doublets which sum up to $8$. By working with these truncations we guarantee a conservation of the   probability, i.e. $\int|\Psi|^2dxdy$ equal to 1, with error smaller than 5 decimal digits at every time $t$.

The steps that we followed for the calculation of Bohmian trajectories were the following:
\begin{enumerate}
\item Construction of the Hamiltonian matrix with respect to the basis $|\Phi_m\rangle$ of the energy eigenstates of the unperturbed system ($\epsilon=0$).
\item Numerical diagonalization of the Hamiltonian matrix for given values of the  parameters.
\item Construction of $\Psi(t)$ as superposition of $|\Phi_m\rangle$. 
\item Calculation of the real and of the imaginary part of $\Psi$, $\Psi_R$ and $\Psi_{Im}$ and then of $|\Psi|^2$.
\item Calculation of the derivatives of $\Psi_R$ and $\Psi_{Im}$ and construction of the Bohmian equations of motion.
\end{enumerate}
All the above steps were  carried out with the Maple 2016 computer algebra system. However, the resulting equations of motion had a very lengthy  and complicated form and thus their integration could not be made with Maple. That is why we passed this step to a hybrid Python 3-Fortran code.

Due to the remarkable stiffness of the Bohmian system we used the  LSODA integrator of adaptive step of the ODESPY module \cite{langtangen2016primer}. By doing so we managed to calculate the Bohmian trajectories with absolute and relative error tolerance $10^{-10}$ and $10^{-9}$ correspondingly but at the cost of very large integration times which increase quickly with the number of the energy levels inside $|\Psi\rangle$.

Consequently, the method used in the present work, although generic and very useful for calculations in the field of Bohmian chaos, requires a large amount of computational resources.

\section{Appendix B: Distinguishing ordered and chaotic trajectories}

{A standard tool for the detection of chaos is the Lyapunov characteristic number $LCN$: if we consider the deviation vector $\xi_{\lambda}$ between two nearby trajectories at time $t=\lambda t_0, \lambda=0, 1, 2,\dots$, then the quantity}
\begin{equation}
\alpha_{\lambda}=\ln\left(\frac{\xi_{\lambda+1}}{\xi_{\lambda}}\right)
\end{equation}
{is the so called `stretching number'. Then the `finite time Lyapunov characteristic number' is }
\begin{equation}\label{chi}
\chi=\frac{1}{\lambda t_0}\sum_{i=1}^{\lambda}\alpha_{\lambda}.
\end{equation}
Finally,
\begin{equation}
LCN=\lim_{\lambda\to \infty}\chi.
\end{equation}
{The $LCN$ saturates at a positive value in the case of the chaotic trajectories while it is zero in the case of ordered trajectories.}

{The calculation of the $LCN$  is made by simultaneous integration of the equations of motion along with their variational equations. It is, in general, a demanding time consuming process for a computer. In fact, in the present paper where we work with large truncations of the wavefunction containing high order Hermite polynomials, the Bohmian equations are them self about 20000 lines long when written in standard Fortran 77 code. Thus the resulting system augmented by the variational equations is extremely large and a long time integration takes approximately from $0.5$ up to $5$ hours for each initial condition. This makes the calculation of the LCN for the thousands of initial conditions of the Born distribution, in a reasonable time, an impossible task.}

{However, in our previous works \cite{tzemos2020ergodicity,tzemos2020chaos,tzemos2021role} we showed that the chaotic Bohmian trajectories have common limiting distributions of points in the configuration space, i.e. they are ergodic (with the exception of partial ergodic trajectories \cite{tzemos2023unstable}). Namely, if we consider a square grid of cells in the configuration space and count how many times the trajectory passes through each cell of the grid then we find that similar distributions, i.e. similar colorplots. This subject was studied in detail in the above works where we used  the Frobenious distance of the underlying matrices of the colorplots in order to compare them quantitatively.}

\begin{figure}
\centering
\includegraphics[scale=0.25]{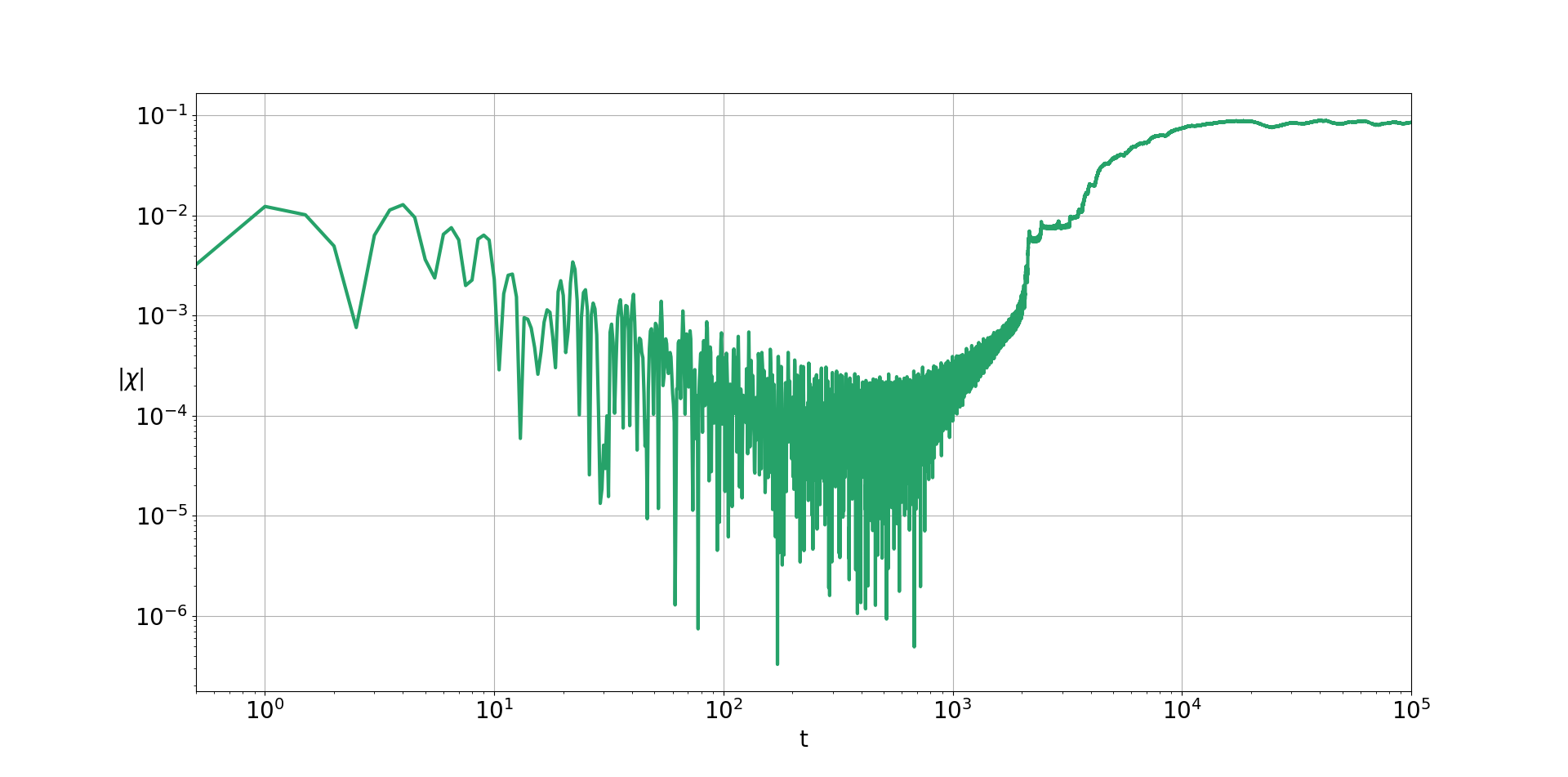}[a]
\includegraphics[scale=0.25]{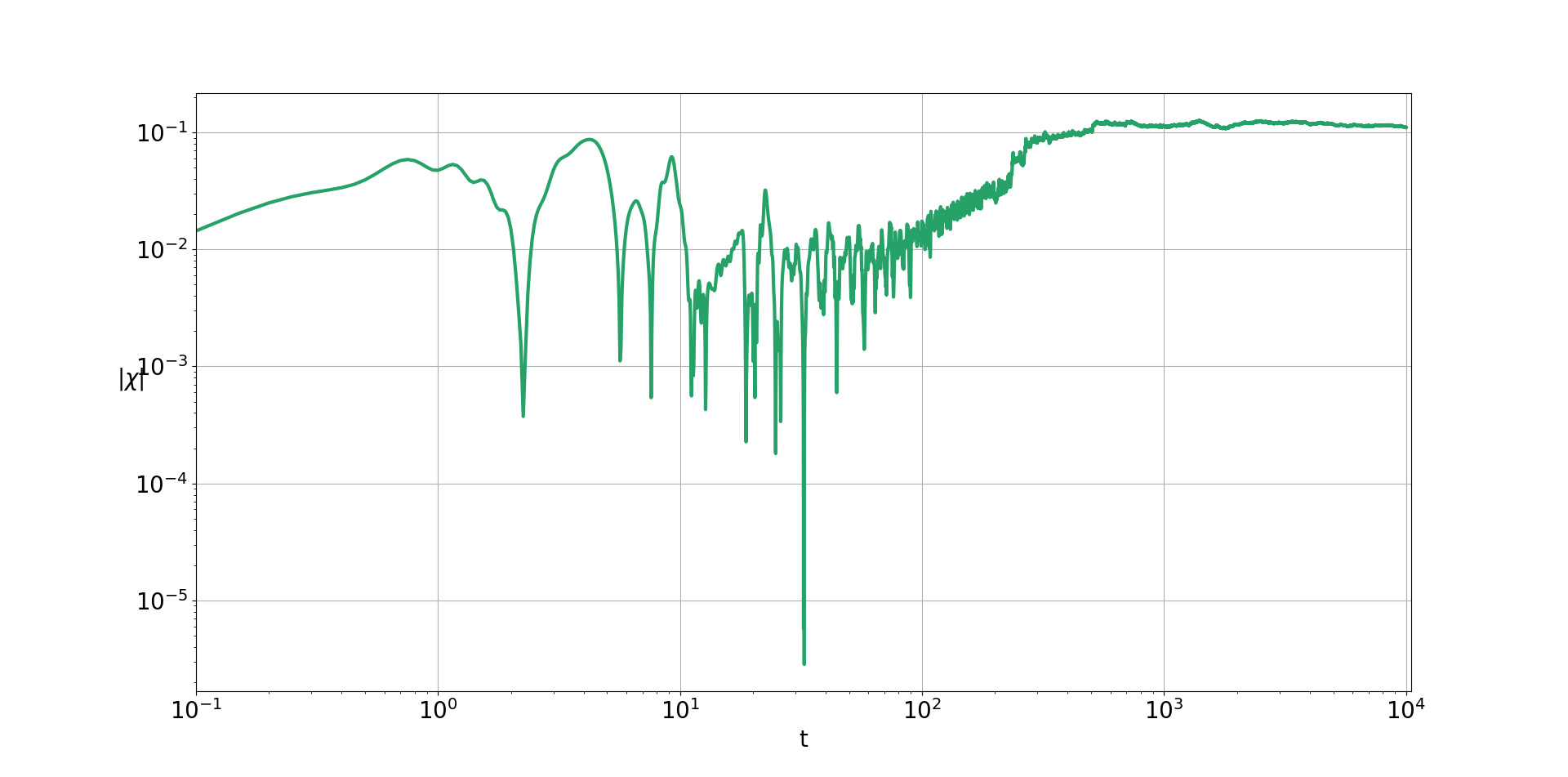}[b]
\caption{The LCNs of Fig.~\ref{traj}a ($\epsilon=0.01$ and Fig.~\ref{traj}b ($\epsilon=0.05$) correspondingly. Both trajectories are chaotic but chaos emerges in shorter times as $\epsilon$ increases.}\label{lcns}
\end{figure}

{In fact we can exploit the ergodicity of the chaotic trajectories in order to avoid the calculation of thousands of LCNs. Namely, we can integrate the Bohmian trajectories up to a very large time and find the corresponding colorplot. The colorplots of the chaotic trajectories cover the largest part of the support of the configuration space (where as support we mean the area where the probability density takes non negligible values), while those of the ordered trajectories are confined and of course not similar for all of them (the depend on the initial condition). Thus whenever the colorplot is close to that of the ergodic picture the trajectory is characterized as chaotic. This practical method has been tested several times in our series of works with the exact calculation of $\chi$ and we never found a deviation. For example in Fig.~\ref{lcns}} {we show the values of $\chi$ (Eq.~\ref{chi}) that approach the Lyapunov characteristic number $LCN$ in the cases of the  trajectories in Figs.~\ref{traj}a,b}.

{We see that in the fist case where the coupling is small the LCN decreases with a power law, a characteristic of ordered trajectories but for $t\simeq 1500$ and above chaos enters, due to the major deformation of $|\Psi|^2$ as we showed graphically in Fig.~\ref{snaps}a. The same hold for the case $\epsilon=0.05$ where chaos enters at a much smaller time $t\simeq 100$ as shown in Fig.~\ref{snaps}b.  We observe also that in the second case the limiting value of the $LCN$ is higher. Thus while $LCN$ is necessary for the exact calculation of the degree of chaos in a trajectory, it is not necessary if we simply want to find approximately the time
beyond which a trajectory exhibits chaotic behaviour.}

\section*{Acknowledgements}
This research was conducted under the program of the Research Committee of the Academy of Athens: ``Study of Order and Chaos in Quantum Dynamical Systems" (200/1026).

\bibliographystyle{elsarticle-num}
\bibliography{bibliography}

\end{document}